\documentclass[10pt]{article}
\usepackage{amsmath,amssymb,amsthm}
\usepackage{epsfig}
\usepackage{graphicx}
\usepackage[numbers,sort&compress]{natbib}
\usepackage{rotating}
\usepackage{url}
\usepackage{xcolor}
\usepackage{color}
\usepackage{latexsym}
\usepackage{slashed}
\usepackage{float}
\usepackage{array}
\usepackage[english]{babel}
\usepackage[T1]{fontenc}
\usepackage[utf8]{inputenc}
\usepackage{booktabs}

\title{Experimental study of breathers and rogue waves generated  by random waves over non-uniform bathymetry}
\author{A. Wang$^1$,  A. Ludu$^{2}$, Z. Zong$^{1,3,4}$, L. Zou$^{5,6,7}$, Y. Pei$^{5}$ 
\\
\\ \small{1. School of Shipbuilding Engineering, Dalian University of Technology}
\\ \small{Dalian 116024, Liaoning, China}
\\ \small{2. Dept. Mathematics \& Wave Lab, Embry-Riddle Aeronautical University}
\\ \small{Daytona Beach, FL, USA}
\\ \small{3. Collaborative Centre of Advanced Ships and Deepwater Engineering, DUT}
\\ \small{Dalian 116024, Liaoning China}
\\ \small{4. Liaoning Deepwater Floating Structure Engineering Technology Lab}
\\ \small{Dalian, Liaoning, China}
\\ \small{5. School of Naval Architecture, Dalian University of Technology}
\\ \small{Dalian 116024, Liaoning, China}
\\ \small{6. State Key Laboratory of Structural Analysis for Industrial Equipment, DUT}
\\ \small{Dalian 116024, Liaoning, China}
\\ \small{6. Collaborative Innovation Center for Advanced Ship and Deep-Sea Exploration}
\\ \small{Shanghai 200240, PR China}
\\}

\begin{document}	
\maketitle 
{\let\thefootnote\relax\footnotetext{{\em Emails}: wam@mail.dlut.edu.cn,  ludua@erau.edu, zongzhi@dlut.edu.cn, lizou@dlut.edu.cn, zoulidut@126.com, ygpei@dlut.edu.cn}}
	
\begin{abstract}

\noindent 
		
Experimental results describing random, uni-directional, long crested, water waves over non-uniform bathymetry confirm the formation of stable coherent wave packages traveling with almost uniform group velocity. The waves are generated with JONSWAP spectrum for various steepness, height and constant period. A set of statistical procedures were applied to the experimental data, including the space and time variation of kurtosis, skewness, BFI, Fourier and moving Fourier spectra, and probability distribution of wave heights. Stable wave packages formed out of the random field and traveling over shoals, valleys and slopes were compared with exact solutions of the NLS equation resulting in good matches and demonstrating that these packages are very similar to deep water breathers solutions, surviving over the non-uniform bathymetry. We also present events of formation of rogue waves over those regions  where the BFI, kurtosis and skewness coefficients have maximal values.  
		
\end{abstract}
\vskip0.7cm

\section{Introduction}
\label{sec.intro}

It is very important to predict with greatest accuracy ocean waves for the safety of ships and offshore structures, especially when operating in rough sea conditions where extreme events could arise.  Ocean extreme waves, also known as rogue waves (RW), occur  without apparent warning and have disastrous impact, mainly because of their large wave heights \cite{2016XXX1,2016XXX2}. These highly destructive phenomena have been observed frequently enough to justify advanced studies. Possible candidates to explain the formation of rogue waves in the ocean are presently under intense
discussion \cite{2012XXX,2016XXX1,2012XXX3,2012XXX4}. This topic attracted recently a great deal of scientific interest not only because of the accurate modeling and prediction of these extremes and similar structures \cite{2016XXX6,2016XXX7,2016XXX8,2016XXX9}, but also because of the interdisciplinary nature of the modulation instability (MI) present in weakly nonlinear waves  \cite{2018XXX11,2016XXX3,2016XXX4,2016XXX5}. Explanations solely based on linear wave dynamical theories (constructive interference of multiple small amplitude waves) cannot grasp the  nonlinear coupling between modes, phenomenon which becomes important when the amplitude of the waves increases.

One of the most cited nonlinear approaches for surface gravity wave propagation is the modulation instability (MI)  \cite{2012XXX5}. Such a phenomenon can be described by the evolution of an unstable wave packet which absorbs energy from neighbor waves and increases its amplitude, reaches a maximum and then transfers its energy back to the other waves \cite{2005XXX}.   

The most common mathematical  model for such unstable modes describing  the nonlinear dynamics of gravity waves is the nonlinear Schr\"{o}dinger equation (NLS) \cite{2018XXX,3,2018XXX10,2018XXX11} or extended versions of it \cite{dysthe,truls96,2011exp}. 
Exact solutions of the NLS equation provide feasible models that were successfully used  to provide deterministic numerical and laboratory prototypes both to reveal novel insights of MI  \cite{2016XXX10} and to describe rogue waves.
The reason for the efficiency of the NLS model is that through its balance between nonlinearity and linear dispersion it can describe well the occurrence of Benjamin-Feir instability, and the associate  nonlinear wave dynamics \cite{2012XXX6,2012XXX7,2012XXX8}. Experimental studies confirmed validity of NLS for deep water waves \cite{2012XXX9,2012XXX10,2012XXX11}. One other advantage of using the NLS is its integrability \cite{2012XXX12}, and analytic form for solutions, especially useful when compared to experimental results.

In NLS models the instability corresponds to various breather solution of this equation \cite{2018XXX10}. The NLS equation (as opposed to the solitons in other integrable nonlinear equations like Korteweg-de Vries KdV) is characterized by a much richer family of coherent structures, namely breather solutions \cite{2005XXX2,2005XXX3,2018XXX,2016XXX,2012XXX,2005XXX,2011exp}. Even if the breathers change their shape during their evolution and hence are not traveling solitons, they maintain their identity against perturbations and collisions. 

Breathers are exact solutions of the nonlinear Schr\"{o}dinger equation (NLS) \cite{2018XXX10,2018XXX11} and describe the dynamics of modulation unstable Stokes waves \cite{2018XXX12} in deep water \cite{2018XXX13,2018XXX15}. The MI starts from an infinitesimal perturbation that initially growths exponentially, and after reaching a highest amplitude decays back in the background wave field \cite{3}.

Such full exact solutions for the NLS equation are given by rational expressions in hyperbolic and trigonometric functions of space and time and are known as Akhmediev breathers (AB) \cite{2018XXX16,2018XXX17} providing space-periodic models to study the Benjamin-Feir instability initiated from a periodic modulation of Stokes waves \cite{2018XXX18,2018XXX19}. 

A limiting situation is the case of an infinite modulation period and corresponding significant double localization. Such solution is described through a rational function called the Peregrine breather (PB)  \cite{2018XXX16,2018XXX17,2018XXX18,2018XXX19,2018XXX20}.  The growth rate of the KM breather is algebraic \cite{3}. Both AB and PB have been considered as possible ocean rogue waves model \cite{2018XXX21,2018XXX22}, and their  features have been investigated experimentally and numerically \cite{2018XXX23,2018XXX24,2018XXX25,2018XXX26}. 

In addition to these MI solutions, NLS equation has also time-periodic solutions in the form of envelope solitons traveling on a finite background, which do not correspond to MI, and are called Kuznetsov-Ma solitons (KM).

It is natural to apply such successful NLS-breathers deep water hydrodynamic models in realistic oceanographic situations where the underlying field is irregular and random \cite{2016XXX16}. Even if initially the  ocean surface dynamics
is narrow-banded, winds, currents and wave breaking may induce strong irregularities. 
Recently it was demonstrated  the possibility of extending NLS models to such broad-banded processes, a fact that becomes valuable in the prediction of extreme events and in extending the range of applicability of coherent structures in ocean engineering. There have been a lot of progress lately in this direction \cite{2016XXX,2005XXX,2018XXX,2011exp}. In \cite{2018XXX37} it is reported the
possibility for exact breather solutions to trigger extreme
events in realistic oceanic conditions. By embedding PB into an irregular ocean configuration with random phases, for example a JONSWAP spectrum
\cite{2016XXX17}, the unstable PB wave packet perturbation initiates the focusing of an extreme event of rogue wave type, in good agreement with NLS and even modified NLS (MNLS) predictions \cite{2018XXX}. In this study rigorous numerical simulations based on the fully nonlinear enhanced spectral boundary integral method shown that weakly nonlinear localized PB-type packets propagate in random seas for a long enough time, within certain range of steepness and spectral bandwidth of the nonlinear dynamical process, somehow in opposition to what the weakly nonlinear theory for narrow-banded wave trains with moderate steepness would predict.

This results are also backed up by recent hydrodynamic laboratory experiments also show that PB breathers persist even under wind forcing \cite{2018XXX36}. From the existing literature, especially the articles published in the last eight years it appears that the role of coherent structures like solitons and breathers in the properties of a system of a large number of random waves is definitely a task of major importance both from fundamental and applications points of view. 

One of the objectives of the present work is to provide a detailed
analysis of our experimental data showing the occurrence out of the random wave field, and survival against nonlinear interactions, and against the effects of traveling over non-uniform bathymetry, of breathers and other coherent modes in a hydrodynamic tank. The second objective is the comparison of these experimental results with exact analytic expressions of PB, AB and KM breathers, and to study the limits of applicability of the NLS equation model for deep water random waves over variable bathymetry.

In the last decade, the propagation of gravity waves over variable
bathymetry profiles has been studied as a possible configuration
enhancing the occurrence of large waves. Different studies have described the statistical properties of gravity waves in this configuration both experimentally and with different numerical methods ranging from KdV models, \cite{25.of.trulsen2012,2018XXX25}, through modified NLS equations,  and Boussinesq models, \cite{trulsen2012}, up to fully nonlinear flow solvers \cite{viotti,ducro}.

Trulsen \textit{et al} shown, \cite{trulsen2012}, that the change of depth can provoke increased likelihood of RW. As waves propagate from deeper to shallower water, linear refraction can transform the waves such that the wavelength becomes shorter, while the amplitude and the steepness become larger, and vice-versa. The dependence of the statistics parameters (spectrum, variance, skewness, kurtosis, BFI, etc.) of long unidirectional waves over flat bottom, versus the depth $h$ is a result of interaction between several competing processes within the nonlinear waves. One one hand, Whitham theory, \cite{with1967}, 
for nonlinear waves predicts that in shallower depth long-crested waves become modulationally stable, hence the modulation instability (MI) tends to decrease with the decreasing of $kh$, and annihilate when $kh< 1.363$ because the coefficient the cubic nonlinear term
vanishes at this threshold. On the other hand 
Zakharov equation (for example \cite{trulsen2012,janono}) predicts increasing of MI through increasing of the waves steepness $\epsilon$ by linear refraction and by static nonlinear $2^{nd}$ order effects with decreasing of the depth. Numerical studies by Janssen \textit{et al}, \cite{jan23}, have shown that shallower water involves the decreasing of kurtosis all together by these effects. 

Nonlinear unidirectional wave fields over non-uniform bathymetry have a different dynamical behavior because the traveling nonlinear waves reach an equilibrium at some depth, and then they loose this equilibrium when running over different depth, and it takes time and space extension for the waves to reach another state of equilibrium.
Numerical studies of NLS solution performed by Janssen \textit{et al}, \cite{24.of.trulsen2012}, show that the combination of focusing and nonlinear effects result in increasing of kurtosis when waves run over shallower depths, for example when $kh: 20 \rightarrow 0.2$. The same strong non-Gaussian deviation towards shallower bottom was confirmed in numerical studies by Sergeeva \& Pelinovsky \textit{et al}, \cite{25.of.trulsen2012}. More interesting though, the change in waves' kurtosis with the depth depends itself on what side of the slope the waves are investigated. In experiments over sloped bottom Trulsen \textit{et al}, \cite{trulsen2012}, shown that waves propagating over a sloping bottom from a deeper to a shallower domain present a local maximum of
kurtosis and skewness close to the shallower side of the slope, and a local maximum
of probability of large wave envelope at the same location, situation which can generate a local maximum of RW formation probability at that point, results backed by  NLS numerical solutions in \cite{5.of.trulsen2012}.

The present paper provides experimental evidence of the above discussed numerical predictions for long crested waves propagating over non-uniform bathymetry and confirms the experimental results obtained by Trulsen \textit{et al} in \cite{trulsen2012}. In addition to these results, and for the first time in literature we present nonlinear waves generated by random fields and propagating over a slope, followed by a submerged shoal, followed by another slope and a final run-up beach. We study the evolution of the spectrum, skewness, kurtosis and other statistical descriptors while the waves pass over this bottom landscape. Moreover, we detected the formation and persistence of coherent wave packets, possibly breathers, traveling over this variable bathymetry with almost constant group velocity and stable evolution and shapes.

The paper is organized as follows: In section 2 we present the experimental setup and the type of waves and their physical parameters we are using, and also how the results were collected and analyzed. We identify traveling stable wave packages in the random wave field. In section 3 we analyze the experimental results with respect to the waves steepness, Ursell number, MI, solitons and RW  conditions of formation like skewness, kurtosis and BFI. We also investigate the possibility of formation of RW and we evaluate the wave spectra and the probability of distribution of wave heights. In section 4 we present the NLS  theoretical formalism for flat bottom and for non-uniform depth and compare the corresponding exact solutions with our experimental results.  In conclusion we can prove that within the given bottom bathymetry, breathers, solitons and rogue waves deep water phenomena are generated out of the random wave background, are stable, and are little perturbed by the bathymetry of our experiments.

\section{Experimental set up. Random wave fields over non-uniform bathymetry}
\label{sec.exp}

\begin{figure}[h]
\includegraphics[scale=.06]{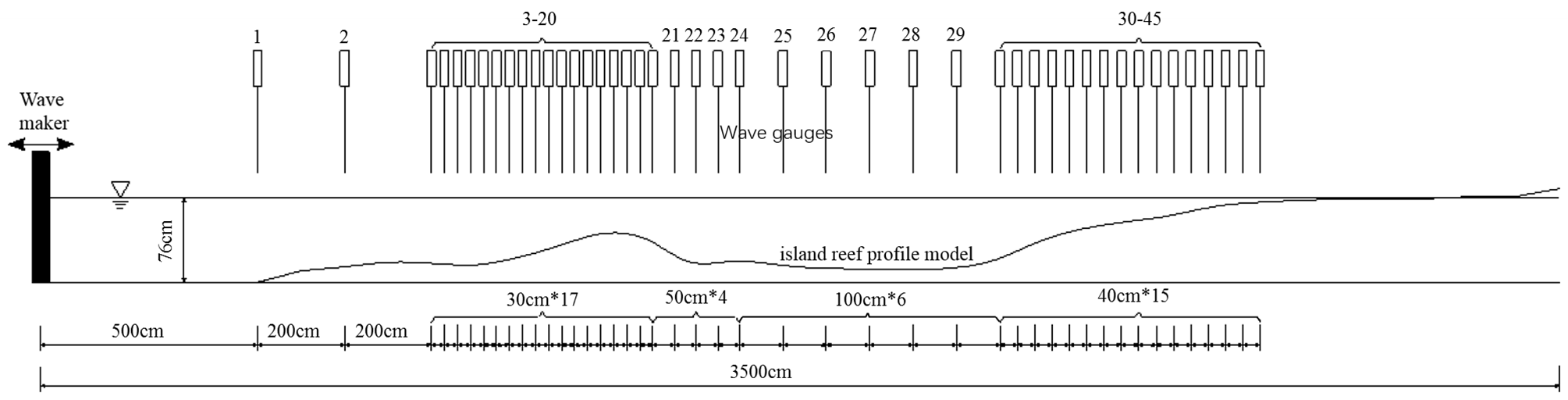}
\centering
\caption{Experiment measurements setup. Positions of the wave gauges with respect to the wave maker and bottom topography.}
\label{fig1}
\end{figure}

\begin{figure}[h]
	\centering
	\begin{tabular}{cc}
		\includegraphics[width=5.4cm,height=4.2cm]{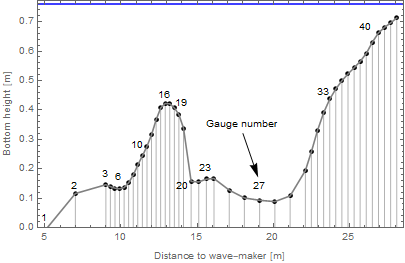} &
		\includegraphics[width=5.4cm,height=4.2cm]{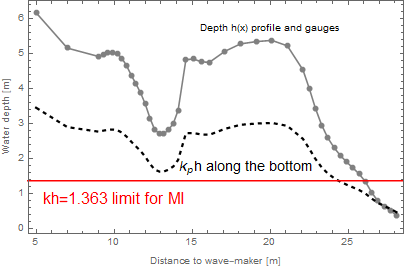} \\
	\end{tabular}
	\caption{\textit{Left}: Bathymetry profile in wave tank, placement of some key gauges and quiescent water level (blue). \textit{Right}: Water depth $h(x)$, expression $k_p (x) h(x)$, and MI extinction threshold $1.363$.}
	\label{tank}
\end{figure}

\begin{figure}[h]
\centering
	\includegraphics[scale=.3]{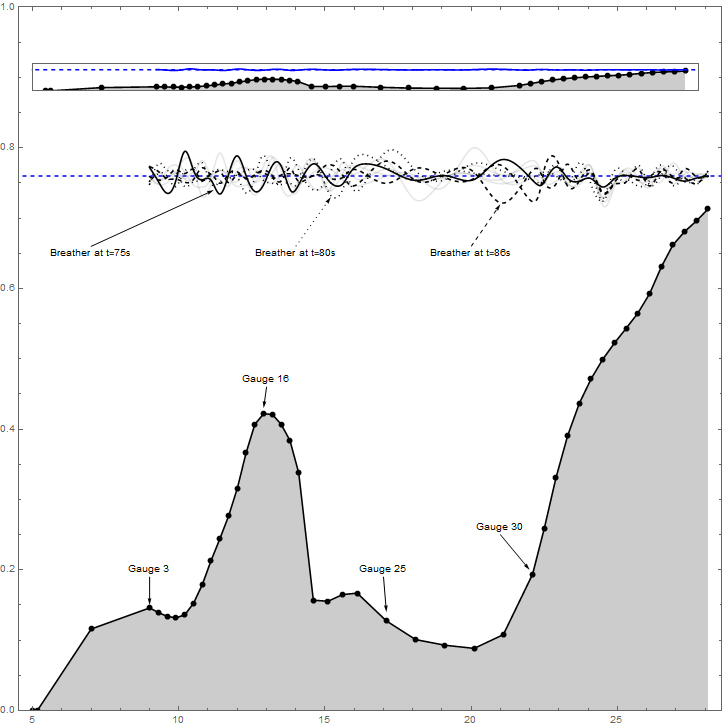} 	
	\caption{Longitudinal section in the wave tank with variable bathymetry. The wave maker is at the left of the frame, the dots represent gauges, and the vertical axis shows depth in meters. The quiescent water level is the dashed blue line and several of our waves are presented to visualize relations between the specific wave heights, wavelengths and depth. We chose moments $t=75$ (solid line), $80$ (dotted line) and $86$s (dashed line) when coherent spontaneous structures (matched with Peregrine breathers) form over gauges number $5\div 6$ (solid line), $24 \div 27$ (dotted line), and $30 \div 34$ (dashed line), respectively. In the upper inset the same picture is present at real scale.}
	\label{fig03}
\end{figure}

\begin{figure}[h]
	\centering
	\includegraphics[scale=.35]{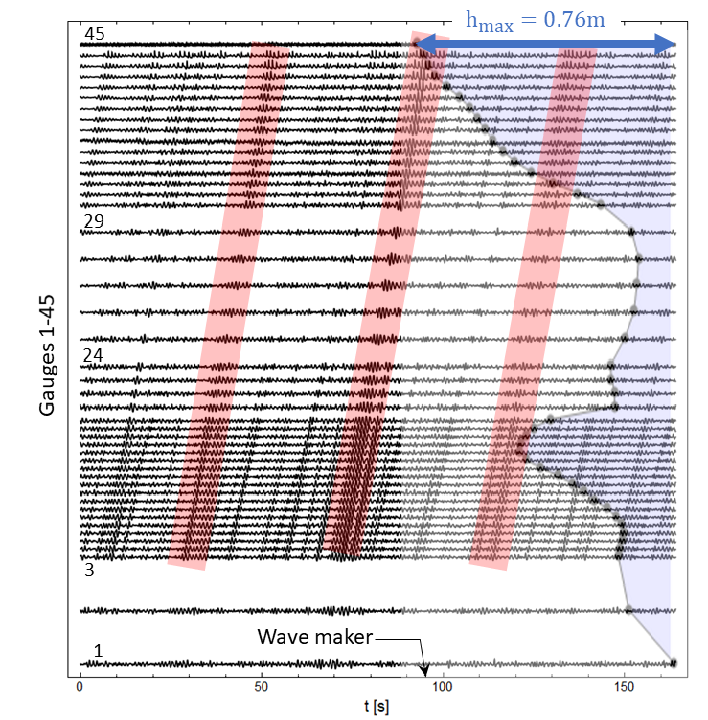} 	
	\caption{Random wave field  of significant wave height  $H_s=3.22$cm, significant period $T_s=0.95$s and variable bottom with depth $h\le 0.76$m. Horizontal axis is time evolution and the 45 gauges signals are lined up along the vertical axis from the wave-maker (bottom) to the run-up (zero water depth on top of the frame). The blue shape represents the bottom profile with the dots being the gauges positions. We identify at least three coherent, stable, and almost uniformly traveling packages, highlighted with red stripes.}
	\label{fig02}
\end{figure}

The experiments have been performed in the wave tank of the State Key Laboratory of Coastal and Offshore Engineering in Dalian University of Technology. The wave tank is  $L_{tank}=50$ m long, $3$ m wide and $1$ m deep. The water tank is provided with a hydraulic servo wave maker at the left end which can generate waves of arbitrary shape with minimum period $0.66$ s at $15$ cm wave maximum height, and an absorbing beach is installed at the other end to avoid wave reflections, Fig. \ref{fig1}. In the present experiment, the bottom has non-uniform shape with the maximum depth of water in the tank  $h=0.76$ m. To insure a unidirectional  wave field and long-crested waves, the wave tank was divided in two sections along its length, of $2$ m and respectively $1$ m each, and we experimented in the wider section.  A number of $45$ resistive wave probes (gauges) were  aligned along the wave propagation direction to measured the wave height.  The surface height of the water at these specified positions is measured with an accuracy of up to 6 significant digits at a sampling $\delta t=0.02$ s, with $170$ s time series length memory. The gauges are placed  as shown in Figs. \ref{fig1}-\ref{fig02}, namely:  first two control gauges, beginning at $9$ m from the wave maker, then  $17$ equidistant gauges with $30$ cm in between, then $4$ gauges at $50$ cm separation, then $6$ gauges separated by $1$ m each, and finally $16$ gauges separated by $40$ m in between. Since the width of the basin is large compared with the characteristic wavelength of our experiments, viscous energy dissipation that occurs mostly on
sidewalls is assumed to be negligible at the center of the basin where our wave probes are located \cite{2011exp}. The bottom shape is inspired by some specific sea floor bathymetry. At the wave maker end the bottom is deep and then gradually increases its heights towards shoal with minimum depth of $h_{min}=0.34$ m at gauge $\#16$, at $x=13$ m from the wave-maker.  From this point the bottom height drops at a larger slope and it reaches its deepest region at $x=20$ m at gauge $\#28$. Then the bottom gradually becomes shallower increasing its height towards a run-up beach all the way to the water surface, see Figs.   \ref{fig1}-\ref{fig02}.

We have carried 40 different experiments by changing the significant wave height $H_s$ and significant period $T_s$, see Table 1, but because of the article length we present here only three relevant cases of $H_s$ and $T_p$.
\begin{table}[h]
	\centering
	\caption{Experimental settings and parameters.}
\begin{tabular}{ccc}
\hline
\hline
Case & $H_s$ (cm) & $T_s$ (s) \\
\hline
B1 & 3.22 & 0.95  1.03  1.13  1.23  1.32  1.41  1.51 \\
B2 & 5.20 & 0.95  1.03  1.13  1.23  1.32  1.41  1.51 \\
B3 & 7.18 & 0.95  1.03  1.13  1.23  1.32  1.41  1.51 \\
B4 & 1.60;   2;   3;   4;   5;   6;   7;   8  &  1.23  \\
 & 5.20;    7.18;    9.20  &0.75 \\
J & 3.22;     5.20;    7.18;    9.20 & 0.85 \\
 & 3.22;     6.20;    8.20;    9.20 & 0.95\\
\hline
\hline 
\label{tab}
\end{tabular}
\end{table}
The sampling time of Cases $B1 \div B4$ is $81.92$ s. In order to study the process of wave evolution in more detail, the sampling time of Cases $J$ is $163.84$ s. A JONSWAP spectrum was chosen for the irregular wave simulation, described by the following parameters \cite{goda1999}
$$
S(f)=\frac{\beta_{J}H_{s}^{2}\gamma^{\delta}}{T_{p}^{4}f^{5}}\hbox{Exp}  \biggl[ -\frac{1.25}{(T_p f)^{4}}  \biggr] ,
$$
\begin{equation}
\delta=\hbox{Exp} \biggl[-\frac{\biggl( \frac{f}{f_p}-1 \biggr)^{2}}{2 \sigma^2} \biggr],
\label{jonswap1}
\end{equation}
with
$$
\beta_{J}\simeq \frac{0.06238}{0.230+0.0336 \gamma -0.185 (1.9+\gamma )^{-1}}\cdot (1.094-0.01915 \ln \gamma),
$$
$$
T_p \simeq \frac{T_{H_s}}{1-0.132 (\gamma+0.2)^{-0.559}},
$$
and the function of wave frequency given by
$$
\sigma= 
\begin{cases}
0.07 & f \le f_p \\0.09 & f> f_p ,
\end{cases}
$$
and were $f_p$ is the spectrum peak frequency, $T_p$ is the spectrum peak frequency, $T_s$ is the significant period, and $\gamma$ is the spectrum peak elevation parameters which we set $\gamma=3.3$.
Given the geometry of the tank and dynamics of the wave maker, ranges of the random waves parameters are limited by three physical constraints: 
deep water condition, \cite{booky}, neglecting capillary waves, and giving the waves enough room to form breathers and eventually rogue waves, that is $\lambda_{capillary}< \lambda <\min \{ L_{tank}, 2 \pi h \}$.  The wave number for the carrier wave  $k_p$ is derived from the linear dispersion relation $k_p =4 \pi^2 /(g T_{p}^{2})$. Under these constraints and according to the parameters chosen in Table 1, the range of peak wavelength that can be excited in the tank becomes $0.85$ m $<\lambda_p < 3.55$ m. 
In our experiments the wavelength and group velocity of the carrier wave changes slightly along the tank because of the non-uniform bottom. In the deep regions at gauges $\#2 \div 7$ and $\#22 \div 29$ ($h=0.65 \div 0.76$ m), or at $x=5-12$ m and $14-22$ m from the wave-maker, see Figs. \ref{tank},\ref{fig03}, we have for the carrier wave period $T_p =0.95$ s deep water, long-waves,  with parameters $\lambda_p =1.407$ m, $k_p =4.49 \hbox{m}^{-1}$ and $v_g=0.76$ m/s. In the intermediate region over the shoal ($h \simeq 0.35$ m) at gauges $\#11 \div 19$, or at $x=12-14$ m from the wave-maker, we still have deep water long-waves with parameters $\lambda_p =1.31$ m, $k_p =4.79 \hbox{m}^{-1}$ and $v_g=0.85$ m/s. Only towards the right end of the beach,  ($h \ge 0.2$ m) at gauges $\#36 \div 45$, or at $x>25$ m from the wave-maker, we  have shallow water  and waves with parameters $\lambda_p =1.13$ m, $k_p =5.55 \hbox{m}^{-1}$ and $v_g=0.89$ m/s.

In the left frame of Figs. \ref{tank} we present the bottom height (topography) and gauges placement. In the right frame we present also with respect to the distance to the wave-maker, the water depth $h$, and the calculated values of $k_p h$ depending on depth and corresponding wavelength for fixed $T_p$. It appears that everywhere along the tank the condition for  developing MI is fulfilled ($k_p h >1.363$ \cite{2016XXX1,2016XXX2,2018XXX11,2012XXX5}), the deep water NLS equation model is valid for the self-focusing regime of solutions, and wave train modulations will experience exponential growth, see for example Figs. \ref{fig2},\ref{figxxx}.

\begin{figure}[h]
	\centering
	\includegraphics[scale=.35]{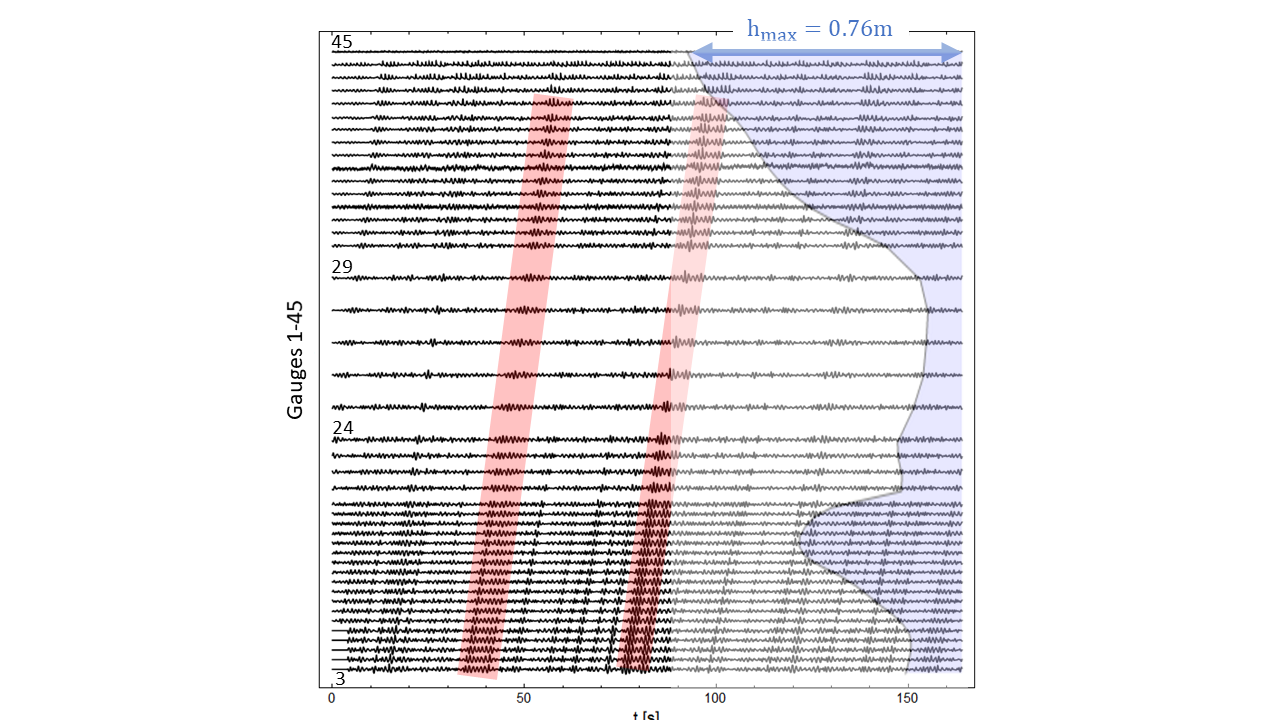} 	
	\caption{Same configuration and parameters as in Fig. \ref{fig02} except here $H_s =5.2$ cm. We still identify at least two coherent, stable  traveling packages, highlighted with red stripes.}
	\label{fig02b}
\end{figure}

\begin{figure}[h]
	\centering
	\includegraphics[scale=.35]{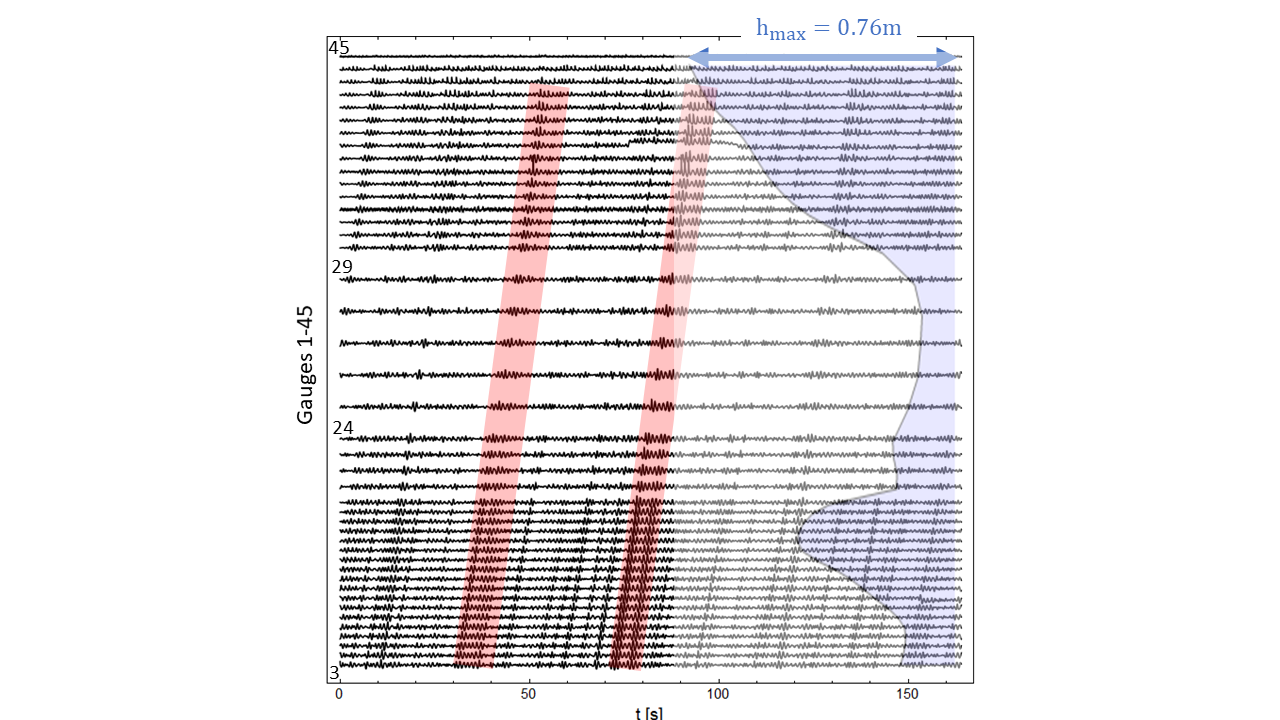} 	
	\caption{Same configuration and parameters as in Fig. \ref{fig02} except here $H_s =6.2$ cm. We still identify at least two coherent, stable  traveling packages, highlighted with red stripes.}
	\label{fig02c}
\end{figure}
\begin{figure}[h]
\centering
\includegraphics[scale=.4]{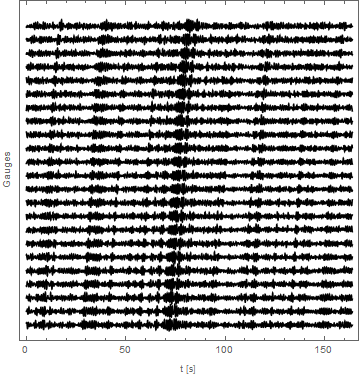} 	
\caption{Gauges $\#3 \div 23$ time series for $T_p =0.9$ s and $H_s =3.22$ cm in the region where gauges are equidistant, but they run over the shoal. The coherent structures, possible breathers, appears traveling with stable shape and group velocity (slope of the line of traveling patterns representing the wave packages) over the variable bathymetry.}
\label{fig2}
\end{figure}
The physical parameters that characterize the evolution of irregular waves are characteristic wave steepness $\epsilon_{p}=k_p H_s /2$, which in our experiments is ranged between $0.015$ and $0.33$, and by  the bandwidth. The spectral bandwidth is determined 
by choosing the peak enhancement factor, which in our case $\gamma=3.3$ induces $\triangle f/f_p =0.095$. The Benjamin–Feir index BFI for the theory of Stokes waves \cite{2016XXX3,2005XXX,sen} which measures the nonlinear and dispersive effects of wave groups is
given by
$$
\hbox{BFI}=\frac{\epsilon_p f_p}{\sqrt{2} \triangle f}.
$$
Beyond a critical value of BFI=$1$ \cite{2011exp} an irregular wave field is expected to be unstable and wave focusing can occur. In our experiments we can cover the range $0.11 < \hbox{BFI} < 2.2$ namely covering all types of sea, from linear waves to stronger MI with development of a rogue sea state, especially since the total length of the measurements covers $28$ m which is larger than the distance over which the MI
is expected to appear \cite{2011exp}. Since the waves in our experiments
may enter occasionally into a strongly nonlinear wave regime, the
NLS equation may not provide a very good fit with these experiments.
\begin{figure}[h]
	\centering
	\includegraphics[width=12cm,height=3cm]{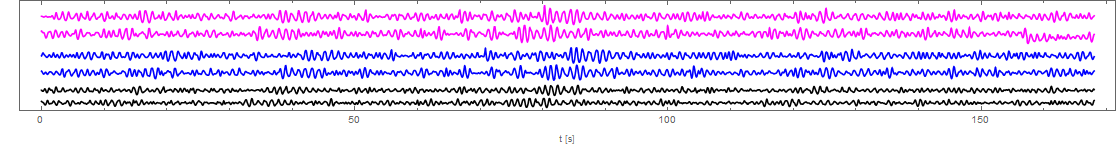}
	\caption{Steepness effect on coherent wave packages. The wave profiles from gauges $\#10$ (at $x=11.1$ m) and $\#20$ (at $x=14.1$ m) for three cases: $H_s=3.22$cm (first two bottom signals, black), $H_s=5.2$ cm (middle two signals, blue) and $H_s=6.2$ cm (upper two, magenta signals) vs. time. Coherent wave packages, most likely Peregrine breathers are spontaneously formed in the random waves and can be observed traveling for as long as $20$ m.}
	\label{figxxx}
\end{figure}

\begin{figure}[h]
	\centering
	\includegraphics[scale=0.42]{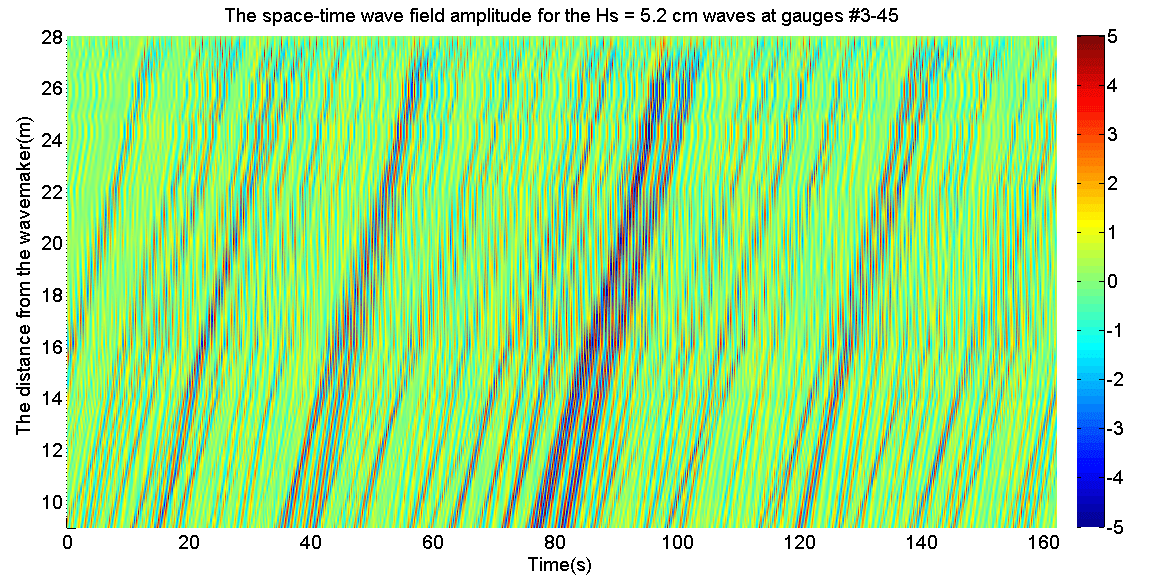}  
	\caption{Density plot of the space-time wave field for $H_s =5.2$ cm waves (wave amplitude scale in cm to the right). The gauges from $\# 3 \div 20$ with $30$ cm between gauges, the gauges from $\# 20 \div 24$ with $50$ cm between gauges, the gauges from $\# 24\div 30$ with $100$ cm between them and the gauges from $\# 30\div 33$ with $40$ cm between them. Higher-order breathers (doublets) can be observed by their red-blue color while propagating uniformly.}
	\label{figYYY}
\end{figure}

\begin{figure}[h]
	\centering
	\begin{tabular}{c}
		\includegraphics[scale=0.4]{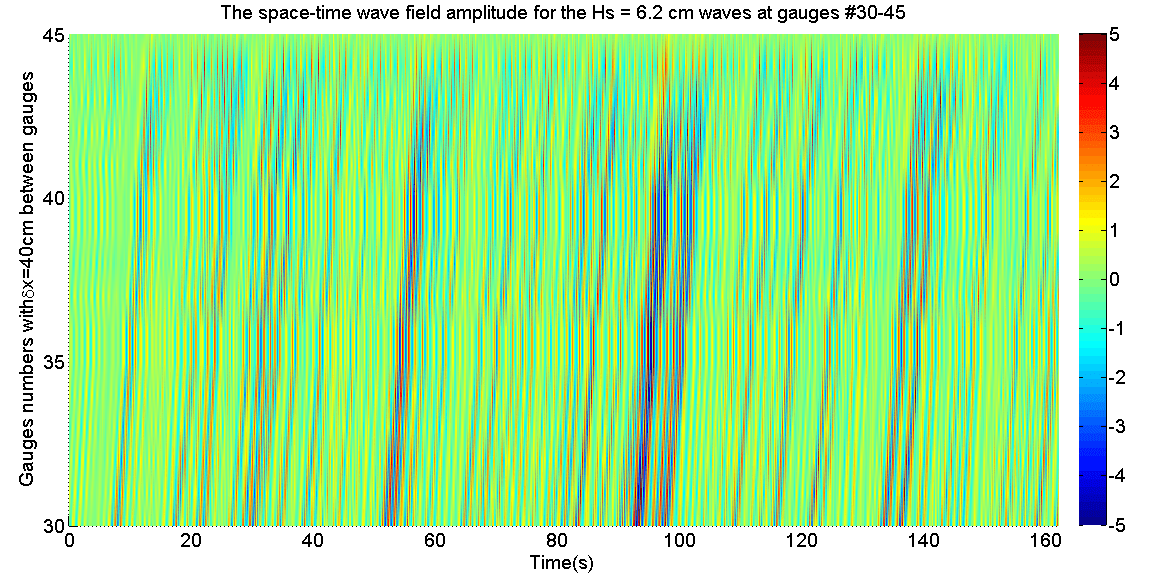} \\
		\includegraphics[scale=0.4]{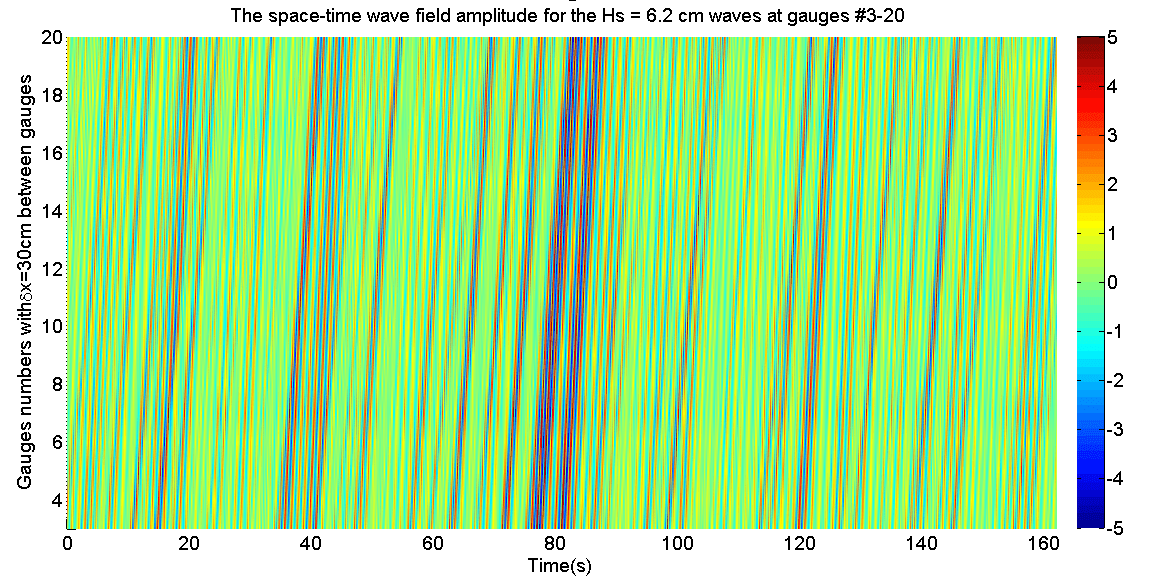}  	
	\end{tabular}
	\caption{Density plot of the space-time wave field for the $H_s = 6.2$ cm waves. Legend for wave amplitude in cm to the right.}
	\label{fig4}
\end{figure}

We first consider irregular JONSWAP waves with significant wave height $H_s=3.22$cm and significant period $T_s=0.95$s over this complex bathymetry. In Fig. \ref{fig03} we present a typical experimental result. In this vertical longitudinal section of the wave tank with variable bathymetry (the gray shape at the bottom) and a wave maker placed at the left of the frame, we show the level of quiescent water by a dashed blue line, on which we overlapped several waves obtained at $t=75$ (solid line showing a nonlinear coherent wave package on top of the shoal), $80$ s (dotted line, showing the same structure which traveled now over the deepest valley) and $t=86$ s (dashed line, when the same coherent package travels up the slope of the run-up). The behavior of the waves shown in Fig. \ref{fig03} is in agreement with the Djorddjevic\'{c}-Redekopp model for deep water with variable bathymetry, using a modified NLS equation with variable coefficients \cite{zaxq}, Eq. \ref{eq.var.bottom.term}. Indeed, in all our experiments the amplitudes and wavelengths of the waves slightly decrease, while $v_g$ slightly increases, over the shoaling region (about gauge $\#16$), and the situation reverses when waves advance over deeper regions (gauges $\#27-29$).

In the upper inset of Fig. \ref{fig03} we present the longitudinal section at real scale, and the same waves, to stress that the all our waves amplitude are negligible compared to the variations in bathymetry.
\begin{figure}[h]
	\centering
	\begin{tabular}{c}
		\includegraphics[width=5cm,height=4cm]{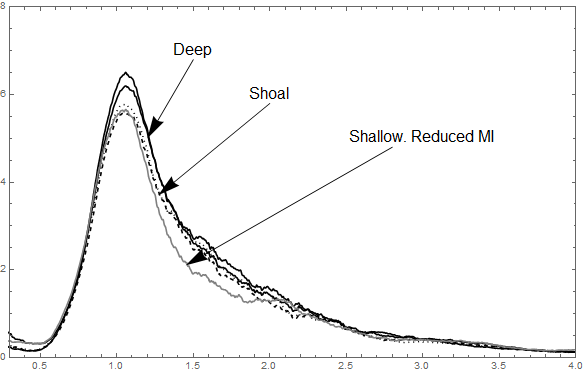} \\
		\includegraphics[width=5cm,height=4cm]{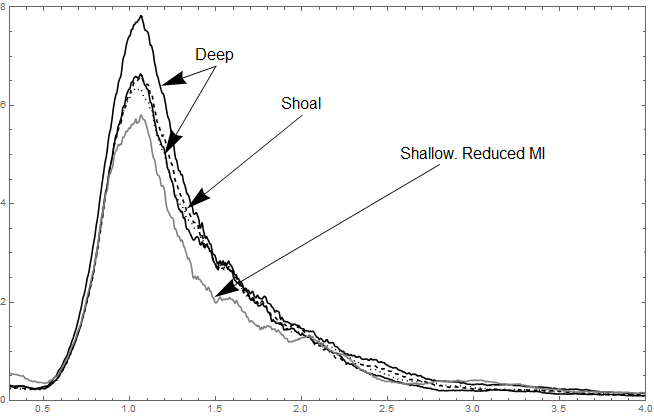} \\
		\includegraphics[width=5cm,height=4cm]{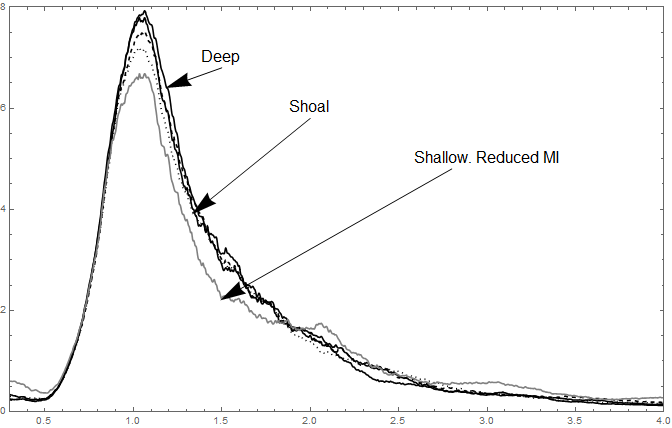} 
	\end{tabular}
	\caption{Linear scale Fourier spectra for $H_s=3.22$ cm (upper frame),  $H_s=5.2$ cm (middle frame), and $H_s=6.2$ cm (bottom frame), all at $T_p =0.95$ s, for  five representative points at gauges: $5, 11, 16, 22$ and $39$.  The spectrum for the deeper sides, before and after the shoal are presented in solid line, the spectra of waves on top of the shoal with dashed line, the spectra at gauges on the ramp about the same height as the shoal by dotted line, and the spectra for regions with $k_p h \ge 1.363$ with gray lines.}
	\label{figfourierglobal}
\end{figure}

\begin{figure}[h]
	\centering
	\begin{tabular}{cc}
		\includegraphics[width=5cm,height=4.5cm]{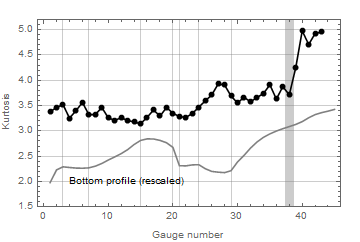} &
		\includegraphics[width=5cm,height=4.5cm]{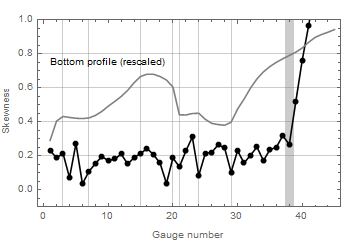} \\
		\includegraphics[width=5cm,height=4.5cm]{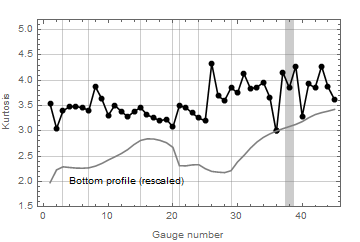} &
		\includegraphics[width=5cm,height=4.5cm]{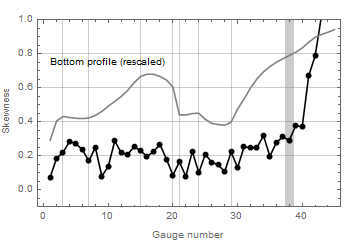} \\ 
		\includegraphics[width=5cm,height=4.5cm]{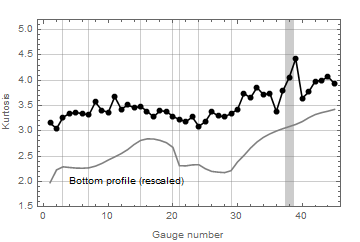} &
		\includegraphics[width=5cm,height=4.5cm]{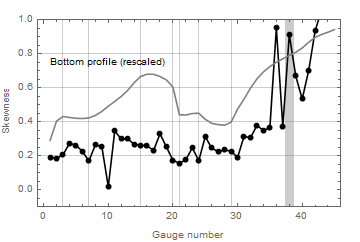} 
	\end{tabular}
	\caption{Kurtosis (left frames) and skewness (right frames) plotted versus the gauge number, next to re-scaled bottom profile (solid line). Upper row represents the waves with  $H_s=3.22$ cm; Middle row represents steeper waves with $H_s=5.2$ cm, and bottom  row represents the steepest waves, with $H_s=6.2$ cm. All have the same $T_p =0.95$ s, and the wave maker is to the left. The vertical grid lines separate different regimes, namely: deep, slope, shoal, quick drop, deep bottom, the deepest, and the run-up beach. The thick gray vertical grid line represents to point $x$ where MI vanishes theoretically, i.e. $k h \rightarrow 1.3.63$.}
	\label{figwkusk}
\end{figure}

For every experiment of generation of random waves we noticed the formation some localized traveling coherent wave packages. These structures, once formed, keep traveling with almost same group velocity over the variable bathymetry, over the shoal and tend to disintegrate when the $kh=1.363$ criterion for MI is not fulfilled anymore, that is around gauge number $39-41$. In Fig. \ref{fig02} we present such an example of a $164$ s long time series (horizontal axis time) as measured by different gauges lined up along the vertical axis. The traveling coherent structures are identified (three of them, for example, are highlighted in red stripes in the figure). These wave packages propagating approximately constant with the peak group velocity of order $v_g \simeq 0.815$ m/s. A larger image for a typical such time series only for gauges $1-23$ is given in Fig. \ref{fig2} where one can detect better the occurrence and stability over the shoal of the nonlinear coherent packages: one begins at $t=27$ s and another larger one begins at $t=68$ s. In Fig. \ref{figxxx} we present in more extended detail wave profiles for $H_s=3.22$cm and $160$s duration time-series measured at 5 locations  (gauges $1, 2, 23, 30$ and $41$) to observe better the nonlinear coherent formations that are spontaneously formed in the random waves and that travel for as long as $20$m.
\begin{figure}[h]
	\centering
	\includegraphics[scale=.36]{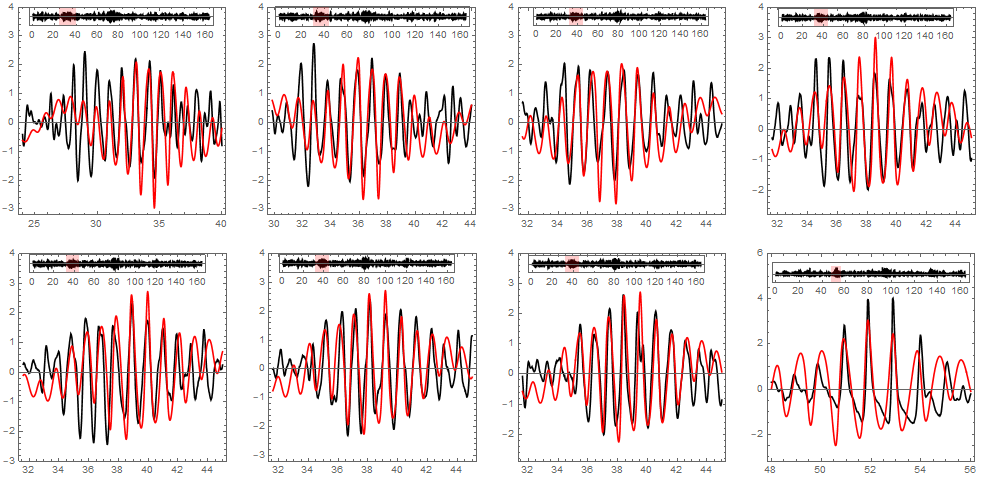} 
	\caption{Black curves in the main frames are the experimental wave profiles for $H_s = 3.22$ cm, measured at gauges $3,13,20,21,22,23,24,26,44$ (from upper left corner clockwise), thus covering a fetch of $21$m beginning at $9$ m from the wave maker. The time interval is shown in the upper inset highlighted in red. Red curves are theoretical KM breather solutions of NLS equation for deep water. The only parameters changing from one frame to another are the origin time, while the  rest of the KM breather parameters ($A_0,\alpha$) are the same for all frames, fact which validates the correctness of our model.}
	\label{figbreath}
\end{figure}

\begin{figure}[h]
	\centering
	\includegraphics[scale=.36]{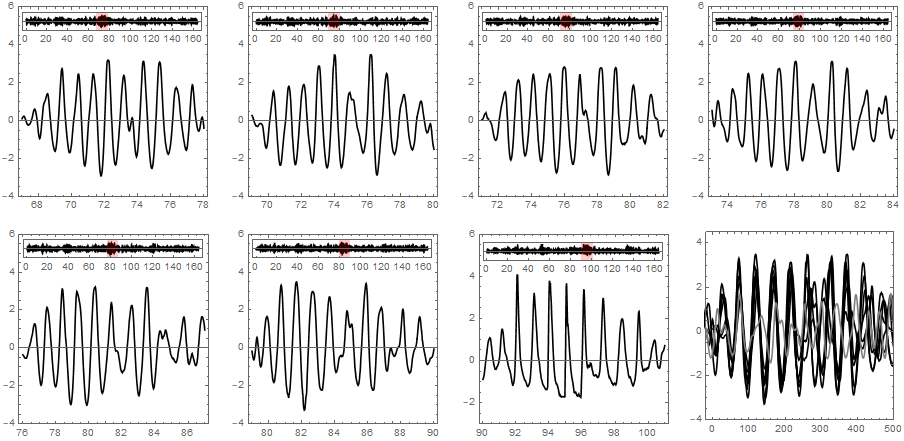} 
	\caption{From left: black curves in the first $7$ frames are experimental wave profiles measured at gauges $3,12,19,20,21,22,23,25,43$, thus covering a fetch of $21$ m beginning at $9$ m from the wave maker. The time interval is shown in the upper inset highlighted in red. The last frame represents an overlap of all these $7$ frames, shifted in time correspondingly. The final frame shows a clear match of the same type of behavior for this coherent traveling group, and the likeliness to be described as a breather, possibly a higher-order breather.}
	\label{figbreathdouble}
\end{figure}

\begin{figure}[h]
	\centering
	\begin{tabular}{cc}
		\includegraphics[width=5cm,height=5cm]{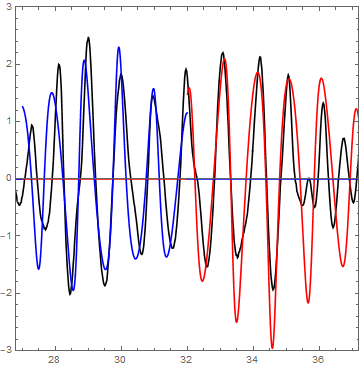} &
		\includegraphics[width=5cm,height=5cm]{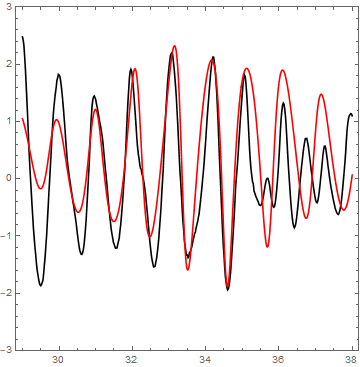} \\
	\end{tabular}
	\caption{Matching $H_s =3.22$ cm waves. Left: gauge $3$ at $t=28-37$ s matched with two KM solitons. Right: gauge $13$ at $t=30-44$ s matched with one KM soliton.}
	\label{fit1}
\end{figure}

\begin{figure}[h]
	\centering
	\begin{tabular}{cccc}
		\includegraphics[width=2.5cm,height=1cm]{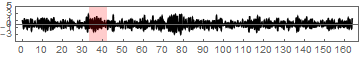} &
		\includegraphics[width=2.5cm,height=1cm]{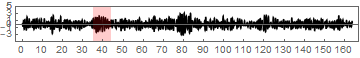} &
		\includegraphics[width=2.5cm,height=1cm]{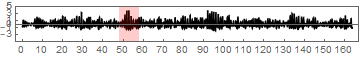} &
		\includegraphics[width=2.5cm,height=1cm]{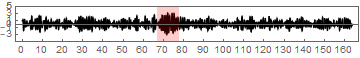} \\	
		\includegraphics[width=2.5cm,height=2.5cm]{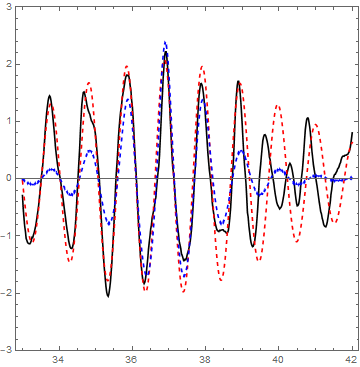} &
		\includegraphics[width=2.5cm,height=2.5cm]{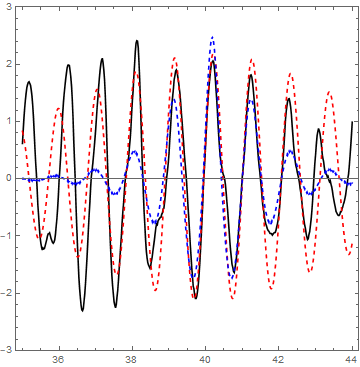} &
		\includegraphics[width=2.5cm,height=2.5cm]{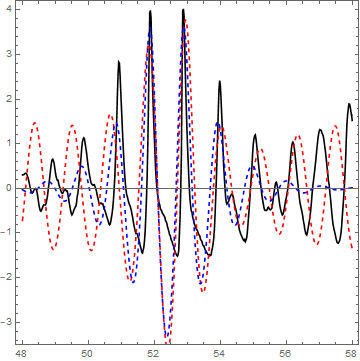} &
		\includegraphics[width=2.5cm,height=2.5cm]{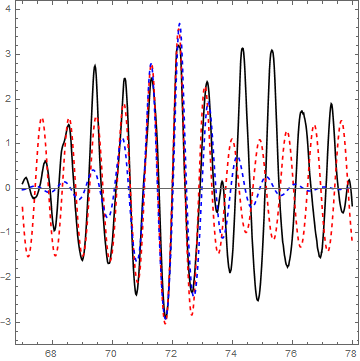} \\	
	\end{tabular}
	\caption{$H_s =3.22$ cm waves. Black curves are experiments, red curves are Peregrine solitons, and blue curves are KM solitons. From left, first three frames represent matching an earlier formed coherent package: gauge $13$ at $t=34-42$ s; gauge $23$ at $t=36-44$ s; gauge $44$ at $t=48-58$ s. Gauge $3$ at $t=68-78$ s matching a later formed coherent package.}
	\label{fit2}
\end{figure}

\begin{figure}[h]
	\centering
	\begin{tabular}{ccc}
		\includegraphics[width=3cm,height=1cm]{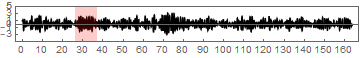} &
		\includegraphics[width=3cm,height=1cm]{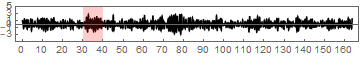} &
		\includegraphics[width=3cm,height=1cm]{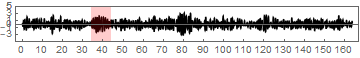} \\	
		\includegraphics[width=3cm,height=3cm]{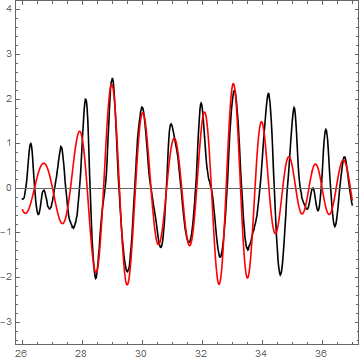} &
		\includegraphics[width=3cm,height=3cm]{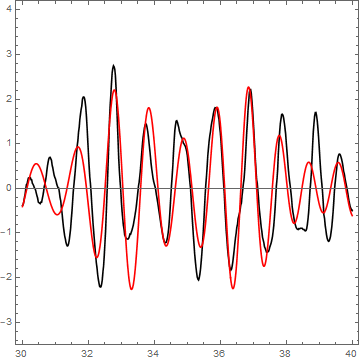} &
		\includegraphics[width=3cm,height=3cm]{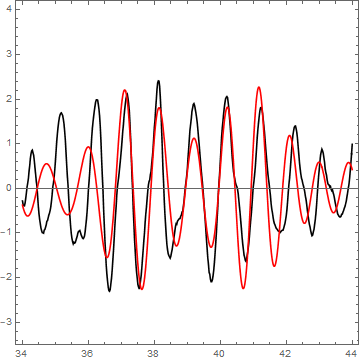} \\	
	\end{tabular}
	\caption{$H_s =3.22$ cm experimental waves plotted with black curves and theoretical match (red curves) with double AB breathers. From left: gauge $3$ at $t=26-36$ s; gauge $13$ at $t=30-40$ s; gauge $23$ at $t=34-44$ s.}
	\label{fit3}
\end{figure}

\begin{figure}[h]
	\centering
	\begin{tabular}{ccc}
		\includegraphics[width=3cm,height=1cm]{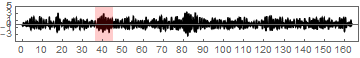} &
		\includegraphics[width=3cm,height=1cm]{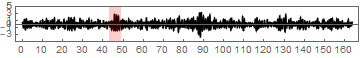} &
		\includegraphics[width=3cm,height=1cm]{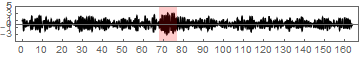} \\
		\includegraphics[width=3cm,height=3cm]{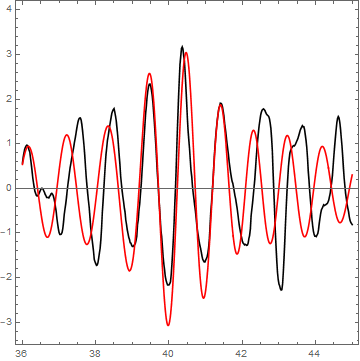} &
		\includegraphics[width=3cm,height=3cm]{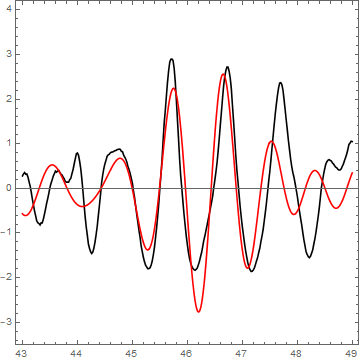} &
		\includegraphics[width=3cm,height=3cm]{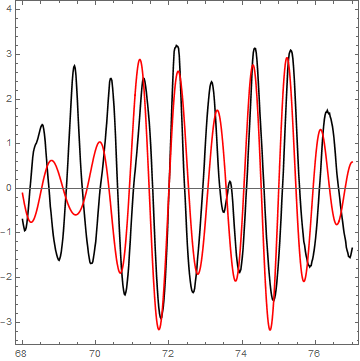} \\	
	\end{tabular}
	\caption{$H_s =3.22$ cm experimental waves plotted with black curves and theoretical match (red curves) with double AB breathers. From left: gauge $26$ at $t=26-46$ s; gauge $31$ at $t=43-49$ s; and again gauge $3$, the latest coherent group at $t=68-77$ s.}
	\label{fit4}
\end{figure}

\section{Analysis of experimental results}
\label{sec.var.bot.stat}

In the analysis of our experiments over variable bathymetry we follow the Trulsen \textit{et al} approach,  \cite{trulsen2012,4.of.trulsen2012,5.of.trulsen2012}, by performing statistics over the time series (and not over space wave field) for the determination of the reference wave and to discern the extreme waves or other coherent structures. In this procedure $H_s =H_s(x)$ becomes  a function of space, and the criteria for identifying breather solutions or RW become local. This approach, supported by numerical studies \cite{5.of.trulsen2012,ducro}, 
allows to isolate the situation favorable for initiation of RW, because linear refraction itself at variable bathymetry points cannot change the probability of RW unless such local criteria for RW are not employed \cite{trulsen2012}. 

Random long-crested waves were propagated over the non-uniform bathymetry. The slope of the bottom topography has values between $1:10$ at the beginning of the shoal (gauges $\#3$, a drop of slope $-1:240$, then slope oscillating between $\pm 1:16$ and finally raising of the beach with slope $1:10 \div 1:35$, all over a length of $28.1$ m. The gauges are placed  as shown in Figs. \ref{fig1}-\ref{fig02}.
Three cases of long-crested random waves were generated with different nominal
significant wave height $H_s$ and constant nominal peak periods $T_p$, as shown in Table 1. The peak wave-number $k_p$ has been computed from the linear waves dispersion relation $\omega^{2}_{p}= g k_p \tanh k_p h$ where $\omega_p =2 \pi / T_p$, $g=9.81$ m$/\hbox{s}^2$.
The characteristic amplitude is calculated as in \cite{trulsen2012} $a_c =H_s / \sqrt{8}$ corresponding to a uniform wave of the same mean power. The Ursell number is $U_r =k_p a_c / (k_p h)^3$.
\begin{table}[h]
	\centering
	\caption{Three significant heights,  at depths: Deep (gauges $3\div 7$, $22 \div 24$, and $30$), Deepest (gauges $26\div 28$), Shoal (gauges $15\div 18$), Beach (gauges $42\div 44$).}
	\begin{tabular}{ccc|cc|cc|cc}
		\toprule
		\toprule
		& \multicolumn{2}{c}{Deep} & \multicolumn{2}{c}{Deepest} & \multicolumn{2}{c}{shoal} & \multicolumn{2}{c}{Beach} \\
		$T_p = 0.95$s  & \multicolumn{2}{c}{$h=0.6$ m} & \multicolumn{2}{c}{$h=0.66$ m} & \multicolumn{2}{c}{$h=0.33$ m} & \multicolumn{2}{c}{$h=0.2$ m} \\
		& \multicolumn{2}{c}{$k_p h=2.7$} & \multicolumn{2}{c}{$k_p h=3$} & \multicolumn{2}{c}{$k_p h=1.6$} & \multicolumn{2}{c}{$k_p h=1.1$} \\
		\cmidrule(r){2-9} 
		$H_s$[cm] &   $k_p a_c$  & $U_r$ &  $k_p a_c$  & $U_r$ &  $k_p a_c$  & $U_r$ &  $k_p a_c$  & $U_r$ \\
		\midrule
		$3.22$ &  $.055$ &  $0.25$ &   $.051$ &  $0.2$ &   $.055$ &   $1.4$  &  $.063$ & $4.6$   \\
		$5.2$  &  $.083$ &  $0.4$ &   $.082$ &  $0.3$ &   $.089$ &  $2.3$ &   $.103$ & $7.5$   \\
		$6.2$  &  $.098$ &  $0.5$ &  $.097$ &  $0.33$ &  $.106$ &  $2.7$ &   $.123$ & $9$  \\
		\bottomrule
		\bottomrule
	\end{tabular}
\end{table}

The three $H_s$ cases for $45$ recording gauges times $8192$ samples taken at  $\delta t=0.02$ s intervals, minus the startup effects provide about $7,000$ peak periods, which, \cite{trulsen2012,dysthe2008}, provide sufficiently reasonable estimates of kurtosis, skewness and overall distribution functions. In Table 2 we present some wave parameters for the three $H_s$ cases investigated, and for four regions of bathymetry called: deep water, the deepest region, shoal and towards the upper parts of the run-up beach, respectively. In all these regions the NLS theory derived by Zakharov, \cite{2018XXX10}, applies and the MI develops in all cases with $k_p h >1.363$ for self-focusing regime. Steepness $\epsilon=k_p a_c $, and  Ursell number was also calculated for all the cases (Table 2) and it shows a very good agreement with the similar cases investigated in \cite{trulsen2012}. $U_r$ has a small value in almost all deep regions, with moderate increased values above the shoal but still in the range of Stokes $3^{rd}$ and NLS equation modeling, and larger values of $U_r$ above the beach where the waves cannot be considered anymore nonlinear deep water, and the character shifts from Stokes $4^{th}$ to $5^{th}$ order to cnoidal behavior, towards breakers in the end of the slope. In shallower water the $2^{nd}$-order nonlinearity of KdV dynamics becomes responsible for the strong correlation observed between skewness, kurtosis and $U_r$, see also Fig. \ref{figbfi2}. 

Some further insight into understanding the waves obtained in these experiments  can be obtained by looking at the wave spectrum at different positions along the wave-tank. The spectra at five representative points for the three cases of significant wave height are shown in Figs. \ref{figfourierglobal} with linear scales. The signal peaks and the Fourier spectra were obtained by using and automatic multiscale peak detection based on the Savitzky-Golay method. All the nominal peaks are centered around the carrier frequency $T_p =0.9s$. There is a slight spectral
development leading to a downshift of the peak, but not very visible, which means that the MI is present almost all over the measurements, except the last few gauges. 
The spectrum corresponding to the deeper sides, independently if this deep region was before and after the shoal are almost identical (solid line in Figs. \ref{figfourierglobal}). The spectra of waves on top of the shoal (dashed line) and the spectra at gauges on the ramp about the same height as the shoal (dotted lines) are not too different from the deep region ones.  However, visible changes in the spectrum show when the waves propagate
towards more shallower regions on the final beach (gray line). At these points, where skewness
and kurtosis attain also maximum values, the spectrum tends to show a second maximum around frequencies doubling the carrier frequency, most likely because of the growth of second order bound harmonics caused by the increased nonlinearity at shallower depth. Further into the shallow region the
spectrum significantly broadens and becomes noisier since energy is shared  to lower and higher frequencies. This situation becomes evident for regions with $k_p h \ge 1.363$ in agreement with the results obtained in \cite{trulsen2012,ducro}.

Nonlinear transfer of energy between modes gives rise to deviations from statistical normality of random waves (Gaussian e.g.). The most convenient statistical properties intended to characterize nonlinear coherent wave packages or extreme wave occurrence are the third and fourth-order moments of the free surface elevation $\eta(x,t)$,
\cite{ducro,trulsen2012}, namely the skewness and the kurtosis defined as
$$
\hbox{Skewness}=\lambda_3=\frac{<\eta-<\eta>^3>}{\sigma^3},
$$
$$
\hbox{Kurtosis}=\lambda_4=\frac{<\eta-<\eta>^4>}{\sigma^4}=\frac{<\eta^4 >}{3 <\eta^2 >^2}-1,
$$
where $<,>$ stands for the average over time and $\sigma$
is the standard deviation of $\eta$, directly related to the significant wave height
$H_s=4 \sigma$. The skewness characterizes the asymmetry of
the distribution with respect to the mean while the kurtosis
measures the importance of the tails. The kurtosis of the wave field is a relevant parameter in the detection of extreme sea states 
\cite{jan23}.

In Figs. \ref{figwkusk} we present the kurtosis of the surface elevation in the left frame and the skewness of the surface elevation in the right frame for the three significant wave heights experimented. The statistical estimates indicate 98\% confidence intervals obtained from $16,500$ selected samples from the original data. For smaller wave amplitude there is a local maximum of kurtosis and skewness, on top of the
shallower edge of the shoal. For larger amplitude waves this kurtosis local maximum shifts towards the beginning of the slope, towards the deeper region. All waves of all heights record a global maximum of the kurtosis in the deepest region, over gauges numbers $24\div 30$, similar to the cases described in \cite{trulsen2012,4.of.trulsen2012,5.of.trulsen2012,24.of.trulsen2012,25.of.trulsen2012}.
This effect is related to the fact that deeper  means $kh$ greater than $1.363$ as seen
in Table 2, and is also related to the spectral evolution leading to a slight downshift of the shallow spectrum with dotted, dashed and gray lines in Figs. \ref{figfourierglobal}. 
For all cases the global maximum of these two statistical quantities is most prominent
at the beginning of the shoal, that is at the positive slope edges of the shoals. 
In all these three cases of different $H_s$ values representing different steepness degrees of the waves, except the end of the run-up beach, the depth is everywhere larger than the threshold value for MI, and not significant shift of the spectral peak can be easily seen. 

Over deep water regions with $kh\ge 1.363$, higher initial BFI (like the waves with $H_s =5.2$cm or $6.2$cm, see the red and blue upper curves in Fig. \ref{figBFIvsSpace}) the kurtosis tends to be  stabilized at higher values as can be seen in the left column in Fig. \ref{figwkusk} for $x=3 \div 8$ m and $x=25\div 30$ m,  for waves with $H_s =3.22$cm,  for $x=7 \div 10$ m and $x=25\div 28$ m for waves with $H_s =5.2$cm, and $H_s =6.2$cm. This result agrees well with previous publications demonstrating that stabilized kurtosis is larger in deep water and smaller in shallower water \cite{5.of.trulsen2012,janono,sen}

However, when $kh\rightarrow 1.363$ beginning at $x \simeq 35$ m, nonlinear effects diminish and the kurtosis decreases towards $3$. This is also visible in Fig. \ref{figwkusk}: for smaller steepness waves with $H_s =3.22$cm kurtosis tends to drop slightly around $x\simeq 36$ m just before the gray vertical stripe in the figure. The dropping effect is more visible at higher steepness, $H_s =5.2$cm
at $36$ m, and again less intense for the steepest waves $H_s =6.2$cm.
After the $1.363$ threshold, the observed oscillations in kurtosis and skewness may be generated by other shallow water effects, Bragg effect, reflection or linear diffraction.  Our results make evident that when a wave field travels over a bottom slope into shallower water, a wake-like
structure may be anticipated on the shallower side for the
skewness and the kurtosis, as it was previously confirmed in \cite{5.of.trulsen2012}. The general expressions for the skewness and the kurtosis of deep water surface evaluated with  Krasitskii’s canonical transformation in the Hamiltonian \cite{jan23}, apply to our cases with $\epsilon \simeq 0.1$ (shallower regions for $H_s =5.2$ cm and all regions for $H_s =6.2$ cm, see Table 2) and our experimental results lineup well with this theory when correlating the values of kurtosis and skewness from Figs. \ref{figwkusk} with the $k_p h$ values from Fig. \ref{tank} (right).

Also, noticing that the value of the BFI decreases with decreasing of the water depth, as we can see it happening for $x>23$ m (or after gauge $33$) in Fig. \ref{figBFIvsSpace}, while the nonlinear coherent structures (which we identified with Peregrine or higher order NLS breathers) keep propagating stable up to shallow water regions, we infer that the probability of RW occurring near the edge of a continental shelf may exhibit a different spatial structure than  for wave fields entering from deep sea and BFI deep water criteria may not apply the same way.

Right after the shoal, both the kurtosis and skewness show oscillations in the values because of a combination between nonlinear effects and linear refraction. One interesting observation resulting from Figs. \ref{figwkusk} is that for small steepness $\epsilon < 0.08$ waves the kurtosis and skewness are larger above the extreme depths $h$ (very deep or $h\simeq h_{max}$, or very shallow or $h\ll h_{max}$), while for larger steepness waves, these two statistical moments tend to acquire their largest values above the sloppy regions of the bottom. This observation can be expressed in a phenomenological relation of the form
$$
\lambda_{3,4} \sim C_1 \biggl(h-\frac{h_{max}}{2} \biggr)^2 + \epsilon C_2 \biggl| \frac{d h}{dx} \biggr|,
$$ 
for some empirically determined constants $C_{1,2}$.
The conclusions obtained from our experimental results and our statistical analysis of kurtosis and skewness coincide with the statistical behavior suggested in the numerical studies from \cite{ducro,4.of.trulsen2012,5.of.trulsen2012,24.of.trulsen2012,25.of.trulsen2012} and with the experimental results obtained for sloped bottom in \cite{trulsen2012}. We have thus shown that as long-crested waves propagate over a shoal and variable bottom in general, local maximum in kurtosis and skewness occur closer to the beginning and the end of the slope, mainly on the shallower side of the slope which can identify these regions as possible locations of high amplitude breathers, multiple breathers and RWs formation.

\subsection{Rogue waves from random background}
\label{sec.stat.breath}

In deep water, long-crested waves are subject to MI, \cite{2016XXX3,3},
which is known to generate conditions for RW formation \cite{2018XXX11,2016XXX1,2018XXX,3,1,dysthe2008,2018XXX13,2018XXX21,2018XXX22,2018XXX24,2016XXX13,sen}. It was also found that nonlinear modulations during the evolution of irregular waves causes spectral development and frequency down-shift, suspected to be related to the occurrence of RW \cite{trulsen2012,socq2005}. In this section we investigate the occurrence of higher amplitude waves, out of the random background, as candidates for RW.
\begin{figure}[h]
	\centering
	\begin{tabular}{cc}
		\includegraphics[width=5.8cm,height=5.5cm]{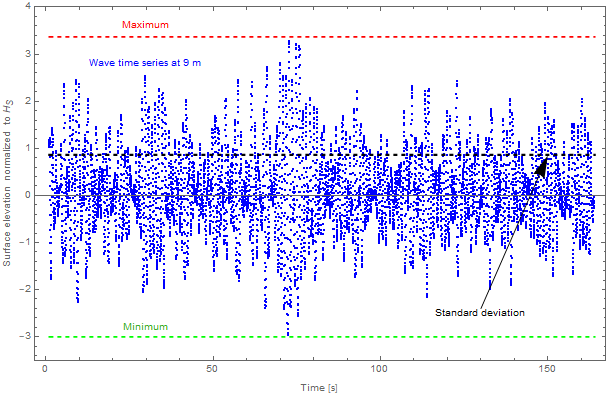} &
		\includegraphics[width=5.8cm,height=5.5cm]{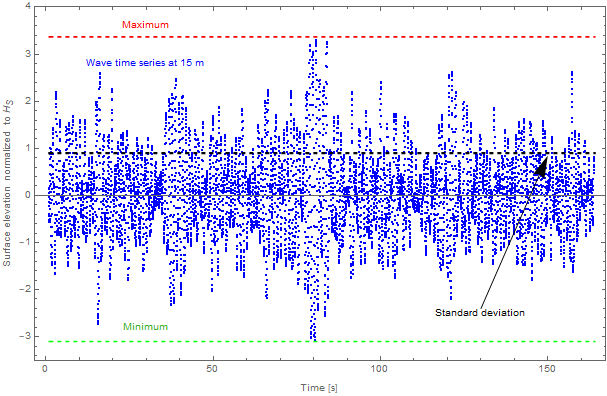} 
	\end{tabular}
	\caption{Time series of the recorded wave amplitudes $\eta(t)$  normalized to their characteristic wave height $H_s =3.22$ cm, measured at $19$ and $26.5$ m (gauges $\# 27, 41$). The horizontal grid lines represent, in the order of their heights: minimum surface (blue), standard deviation (black), and maximum wave (red). The maximum height recorded  is  $2.1 \div 2.42 H_s$, which qualifies them as RWs.}
	\label{fig2exp}
\end{figure}
Extreme height waves, isolated in time and space from the typical background reference wave field is considered to be a Rogue Wave (RW) if it satisfies some common criteria like  $\eta / H_s > 1.25$ or $H / H_s > 2$, where  $\eta$ is the crest elevation, $H$ is the wave height, and $H_s$ is the significant wave height, defined as four times the standard deviation of the surface elevation, \cite{trulsen2012,dysthe2008}. In Fig. \ref{fig2exp} we present an example of two time series recorded in our experiments at two locations, $19$ m (left frame) and $26.5$ m (right frame) from the wave maker. The vertical axis shows the wave amplitude $\eta(t)$  normalized to the characteristic wave height  $H_s =3.22$ cm. We recorded such large amplitudes at  $19, 22, 23.5$ and $26.5$ m (gauges $\# 27, 30, 33, 41$, respectively) from the wave maker. The horizontal grid lines represent, in the order of their heights: minimum surface (blue), standard deviation (black), and maximum wave (red). The maximum height recorded  is  $2.1 \div 2.42 H_s$, which qualifies them as RWs. These events happen over the deep water parts, at the locations where kurtosis and skewness has also local maxima, Figs. \ref{figwkusk}.
\begin{figure}[h]
	\centering	
	\begin{tabular}{cc}
		\includegraphics[width=5.5cm,height=7cm]{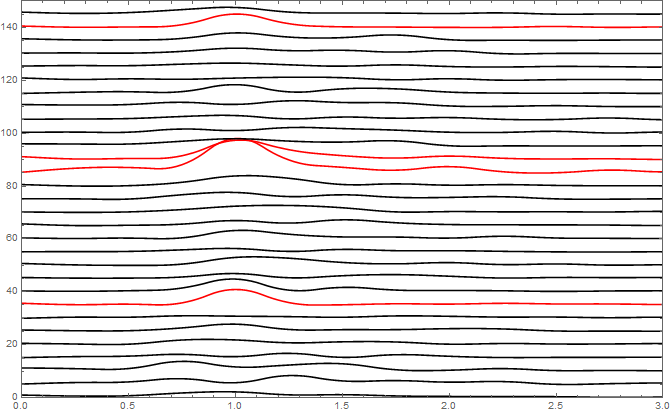} &
		\includegraphics[width=5.5cm,height=7cm]{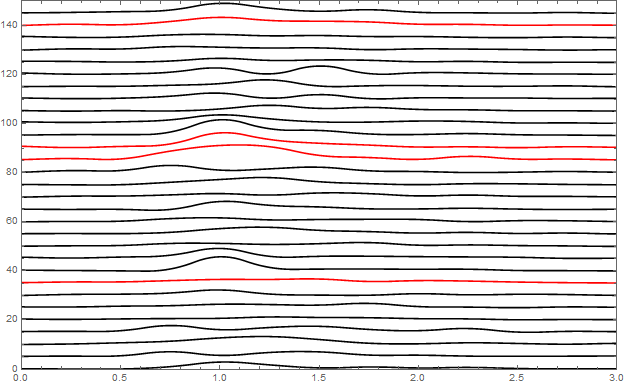} \\
		\includegraphics[width=5.5cm,height=7cm]{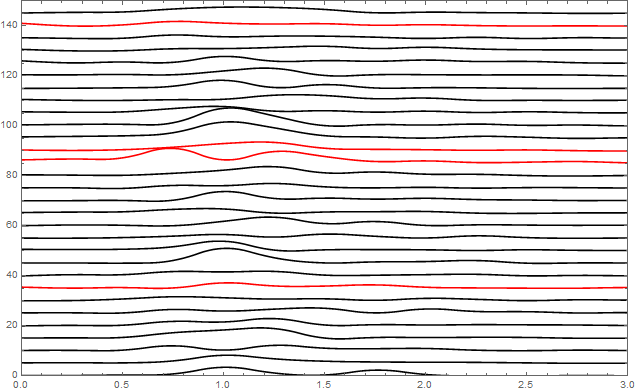} &
		\includegraphics[width=5.5cm,height=7cm]{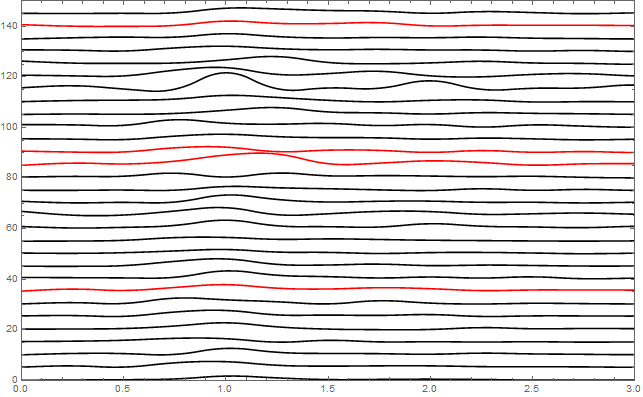}
	\end{tabular}
	\caption{Moving Fourier spectra in time-frequency domain with $4$ s window, for $150$ s long experiments at  $H_s=3.22$cm. Fourier curve's zero-axes are ordered vertically with respect to time, in seconds; but each Fourier spectrum's curve has arbitrary scaling. Horizontal axis represents frequency in Hz. From left upper corner clock-wise the frames represent spectra of waves measured at gauge 3 (at $9$m), 10 (at $12$m), 23 (at $14.1$m) and 41 (at $22.1$m). The red curves represent narrow band width spectra measured at points where coherent packages were detected and assimilated with breathers/solitons/RW, while the black curves represent the wide spectra of random waves filling the background between the breathers.}
	\label{figDFT}
\end{figure}

\begin{figure}[h]
	\centering	
	\begin{tabular}{cc}
		\includegraphics[width=5.5cm,height=7cm]{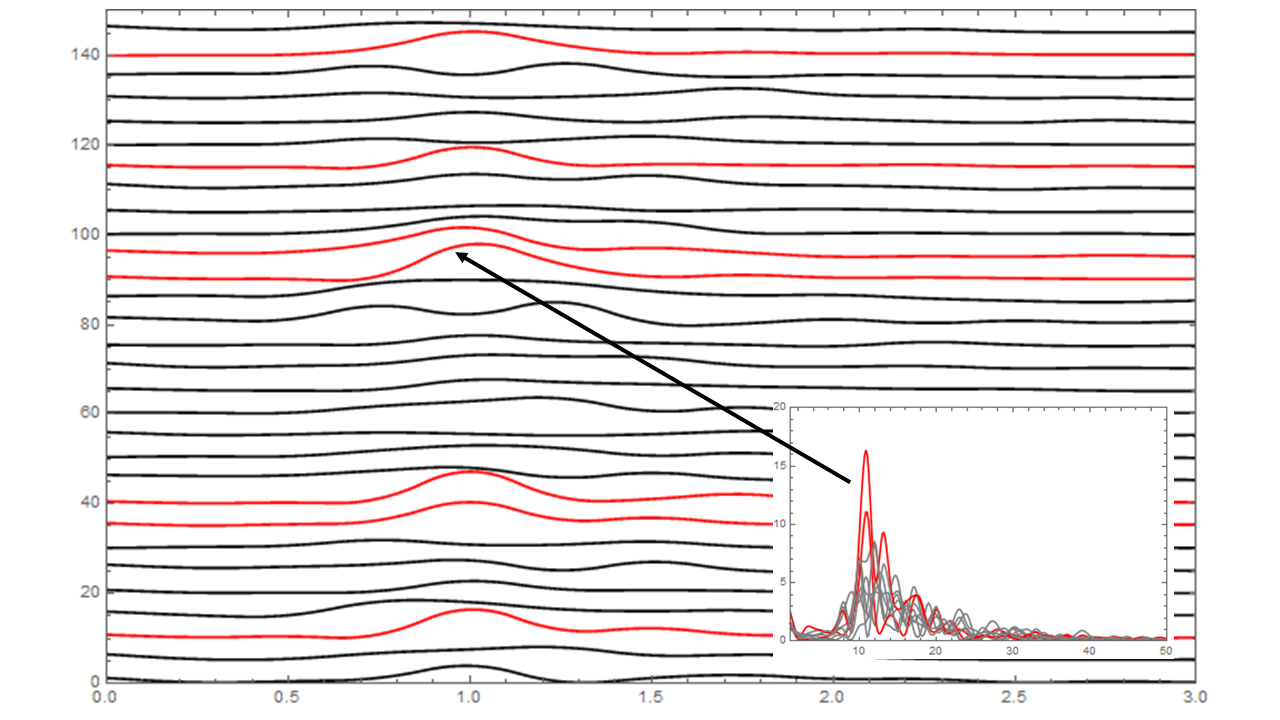} &
		\includegraphics[width=5.5cm,height=7cm]{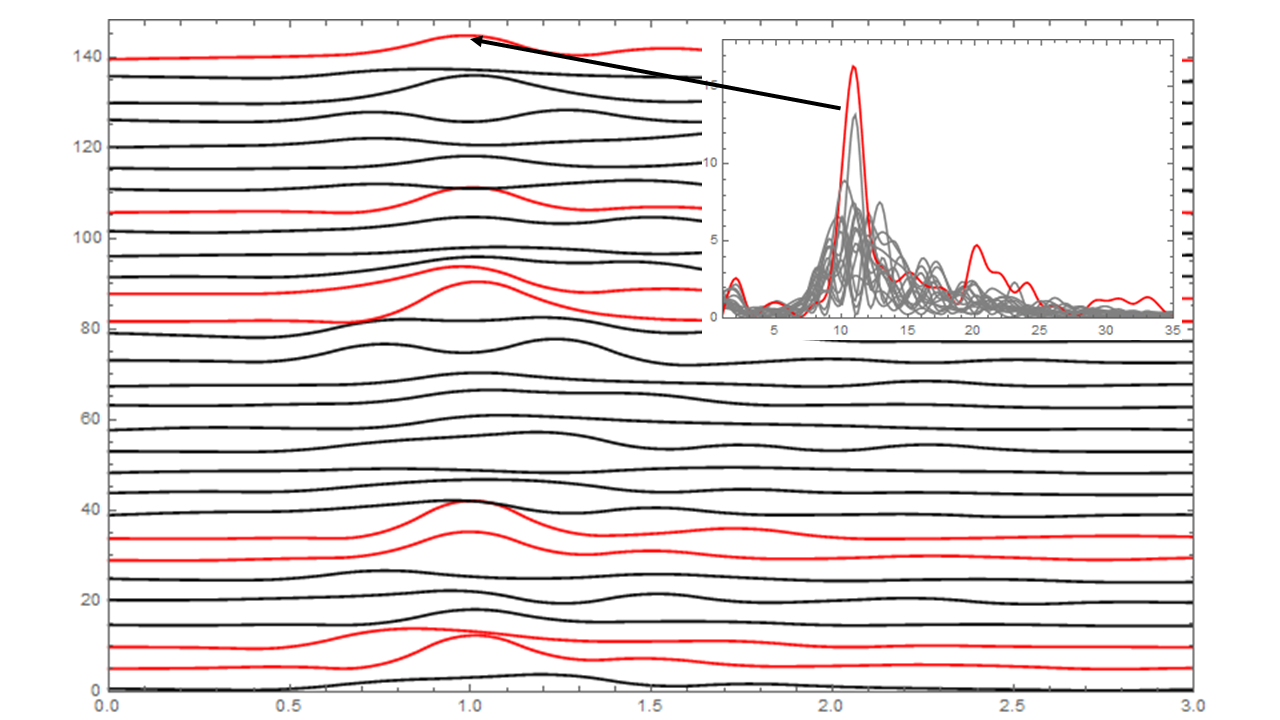} \\
	\end{tabular}
	\caption{Same type of spectral representation as in Figs. \ref{figDFT} for $H_s =5.2$ cm (left frame) and $H_s =6.2$ cm (right frames). In addition, we present in the insets details of the Fourier spectrum of five wave series centered at the moment of time indicated by the arrow: the red spectra are associated to coherent packages identified as breathers/solitons, the gray spectra are the random background waves at nearby points and neighbor moments.}
	\label{figDFT2}
\end{figure}
For a system composed of a large number of independent waves, like the random generation, the surface elevation is
expected to be described by a Gaussian probability density
function. Under this hypothesis, Longuet-Higgins \cite{2005XXX,2005XXX2,2005XXX3}, showed
that, if the wave spectrum is narrow banded, then the probability
probability distribution of crest-to-trough wave heights is given
by the Rayleigh distribution. The distribution was found to
agree well with many field observations \cite{2005XXX}.  Nevertheless, recently \cite{2005XXX2,2005XXX3} it was shown numerically and theoretically that if the ratio between the wave steepness and the
spectral bandwidth this ratio is known as the Benjamin–Feir index (BFI) is large, a departure from the Rayleigh distribution is observed. This departure from the Rayleigh distribution was attributed to the MI mechanism. Moreover, from numerical simulations of the NLS equation it was found \cite{2005XXX} that, as a result of the MI, oscillating coherent structures may be excited
from random spectra. In our experiments we obtained a very good correlation between the waves at regions and during time intervals producing a narrower width spectrum and the corresponding detection (at the same locations and moments of time) of coherent stable, traveling structures, most likely NLS breathers (AB, KM, Peregrine of higher-order breathers, section \ref{sec.theor}). 
\begin{figure}[h]
	\centering
	\includegraphics[scale=.7]{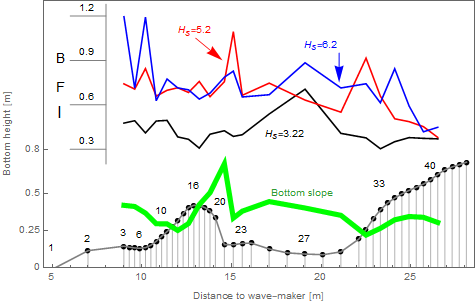}
	\caption{Plot of the average BFI values over $16$s duration, versus space, along the wave tank. Legend: $H_s =3.22$cm (black), $H_s =5.2$cm (red), and $H_s =6.2$cm (blue) for the upper curves. The bottom profile and some gauge numbers are presented by the lowest gray curve, and the slope of the water depth ($dh/dx$) by the green curve. The MI threshold $kh=1.363$ happens around gauges $33-34$, at $x \simeq 24m$ from the wave-maker.}
	\label{figBFIvsSpace}
\end{figure}
In Figs.  \ref{figDFT}, \ref{figDFT2} we present  examples of  Fourier spectra in the time-frequency domain calculated with a $4$ s moving window, at different locations and different moments of time. In these figures, the red curves represent narrower bandwidth wave spectra measured at points where also the coherent packages were detected and assimilated with breathers/solitons/RW, wave packages described in previous sections. Namely, the red curves in Figs.  \ref{figDFT}, \ref{figDFT2} coincide with a good coefficient of correlation ($c=0.76$ Pearson correlation) with the structures highlighted with red stripes in Figs. \ref{fig02}, \ref{fig02b}, \ref{fig02c},  with the coherent packages in uniform motion identified in the mapping of Figs. \ref{fig2}, \ref{figxxx}, \ref{figYYY}, and \ref{fig4}, also they coincide with the packages chosen for theoretical match with breathers and shown in Figs. \ref{figbreath}, \ref{fit1}, \ref{fit2}, and \ref{fit3}, and they are close neighbor  with the extreme amplitude waves shown in Fig. \ref{fig2exp}.
\begin{figure}[h]
	\centering
	\includegraphics[scale=.9]{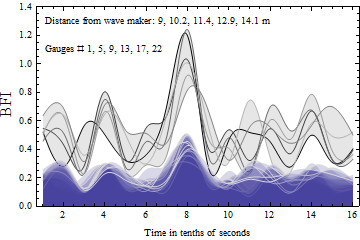}
	\caption{The gray upper curves represent the BFI versus time for 6 selected gauges. The large central peak of BFI$>1$ coincides with the formation of  breathers at that position/moment. All five curves show the same reproducible behavior. The blue profiles at the bottom represent the relative value of the peak frequency in the time series recorded at the 6 selected gauges ($1,5,9,13,17,22$).}
	\label{figbfi1}
\end{figure}

\begin{figure}[h]
	\centering
	\includegraphics[scale=.9]{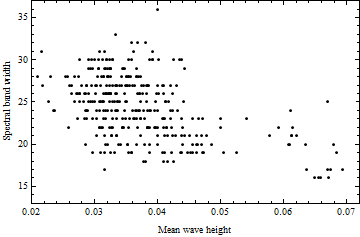}
	\caption{Correlation between the spectral band width (vertical axis in mHz) and mean wave height (in meters) measured at each gauge from $1$ to $22$, for all $H_s$. The points represent a set of 15 mean values of wave spectra and wave heights are evaluated across $150$s time series in samples of $10$s each, for $22$ gauges. The two resulting clusters describe random waves (low wave height, higher frequencies) and breathers (higher waves, lower frequencies).}
	\label{figbfi2}
\end{figure}

\begin{figure}[h]
	\centering	
	\begin{tabular}{ccc}
		\includegraphics[width=3.5cm,height=3.5cm]{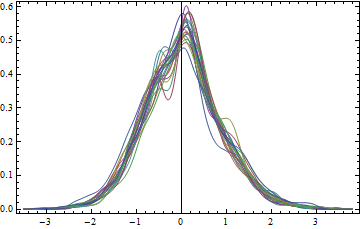} &
		\includegraphics[width=3.5cm,height=3.5cm]{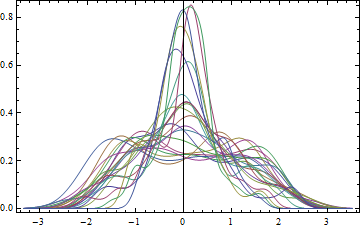} &
		\includegraphics[width=3.5cm,height=3.5cm]{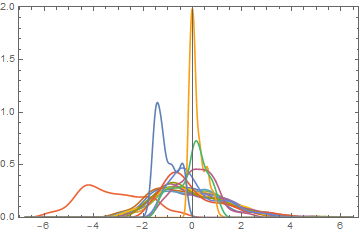}
	\end{tabular}
	\caption{Probability distributions for the wave heights for $H_s =5.2$ cm. Left: mean values calculated across $160$ s time series and $10.1$ m fetch for gauges $1-22$. Center: mean values calculated for the interval $30-36$ s, $5$ m fetch, gauges $1-22$. We note the cluster of narrow band-width spectra associated to the breathers present within this time interval and location. Right: mean values for the interval $88-98$ s, $7$ m fetch, gauges $30-46$ at $22 \div 29$ m from the wave maker, respectively. This spectrum contains mainly unstable structures resembling peakons, and breaking waves.}
	\label{figz1z2z3}
\end{figure}

\begin{figure}[h]
	\centering
		\includegraphics[width=10cm,height=4cm]{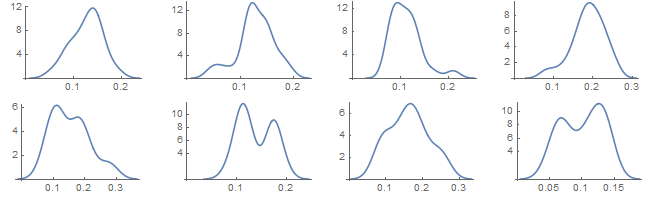}
	\caption{Wave height probability distributions for different moments of time, from upper left corner CW: $t=1, 2, 15, 20, 35, 50, 60$ and $75$ s. Each distribution calculated over $2$ s interval ($100$ samples) over the fetch $9-14$ m (gauges $3-22$) for $H_s =5.2$ cm. Three main modes are present: dominant low amplitude waves at $t=15$ and $35$ s), dominant high amplitude waves at $t=1, 2, 20$ and $60$ s, and  flat PDF distribution, at $t=35$ s. Occasionally the distribution becomes bi-modal. Also we note a cyclic behavior since certain types of PDF tend to repeat.}
	\label{figz4}
\end{figure}

These positive correlations represent an evidence that  MI process takes place in our experimental real long-crested water waves, with high values for the BFI index (the ratio between the wave steepness and the spectral bandwidth) at various depths, on the top of the shoal and equally in the deep regions around the shoal. In the case of our random waves the large values for BFI and the narrower width of the spectra lead to  MI evolution and to a ``rogue sea'' state, that is a highly intermittent sea state characterized by a high density of unstable modes, see Fig. \ref{figbfi1}. Our results are very similar with the same types of studies reported \cite{2005XXX}.

By using the calculations of the spectral bandwidths for all our experimental time series, at different locations and for the three types of significant wave height (steepness), we can correlate these data with the mean wave height. The result is presented in Fig.  \ref{figbfi2}. We notice the formation of two separate clusters of higher positive correlation: one for small waves with large spectral band width, and one more localized for the breather/soliton/RW events described by large wave heights and narrower spectra. 

Another statistical feature which can confirm the formation of coherent traveling packages of breather/soliton types (KM, Peregrine and AB solutions) is the distribution of the probability for the wave heights, which we present in Fig. \ref{figz1z2z3}. The middle frame, representing regions with coherent package formation shows evidence of a cluster of narrow band-width spectra associated to these breathers. In Figs. \ref{figz4}  we present the wave height probability distributions for different moments of time  over a $5$ m length. We observe the formation of three main modes: a dominant low-amplitude mode, a dominant high-amplitude mode, and a flat probability distribution which occasionally tends to shift into a bi-modal unstable mode as predicted by the Soares model \cite{bimodal}.

Our experimental results, mainly gathered in Figs.  \ref{figfourierglobal}, \ref{figwkusk}, \ref{figDFT}-\ref{figbfi1} and 
\ref{figz1z2z3}, \ref{figz4},  are in good agreement with the numerical calculation obtained by Trulsen \textit{et al} (\cite{trulsen2012}a), from the Boussinesq model with improved linear dispersion, and with the experiments presented in Gramstad \textit{et al} (\cite{trulsen2012}b).
Indeed, a significantly increased BFI value, and consequently increase in the probability of RW occurs as waves propagates into shallower water. For smaller $H_s$ and $\epsilon= H_s k_p$ the maximum is smaller and delayed, while for larger steepness the maximum occurs earlier and is larger, Fig. \ref{figBFIvsSpace}. Increased values of skewness, kurtosis, and BFI are found on the shallower side of a bottom slope, with a maximum close to, or slightly after the end of the slope  Figs. \ref{figwkusk}, \ref{figBFIvsSpace}. Maxima of the statistical parameters are also observed where the uphill slope is immediately followed by a downhill slope. In the case that waves propagate over a slope from shallower to deeper water, in the theoretical evaluations from \cite{trulsen2012} it was not found on increase in RW  wave occurrence where the wave parameters were $a k_p= 0.038, a/h = 0.035$, and $Ur = 0.031$. In our experiments, however, we noticed this increase in the BFI, kurtosis and steepness when traveling into deeper, probably because our waves parameter, shown in Table 2, are different: $a k_p > 0.05, \  a/h = 0.04$, and $U_r > 0.2$.

\section{Comparison  with exact solutions}
\label{sec.theor}
	
In this section we present some current theoretical models that can fit our experiments with random waves generated in a $L=50$m long, $2$ m wide wave tank with variable bottom and maximum depth  $h_{max}=0.76$ m present by the wave-maker and at two-thirds of the length, see Figs. \ref{fig1}, \ref{fig03}. Since in all experiments described in section \ref{sec.exp} we notice the formation of stable, traveling coherent wave packages, we present in the subsequent section the match between these waves and deep water breathers. We divide this section in two parts: in the first part we present the corresponding theoretical results for uniform bottom, and in the second part we extend this case to variable bathymetry. 

In the uniform bottom case, for an ideal (incompressible and inviscid) liquid under the hypothesis of irrotational flow, the dynamics of a free surface flow is described by the Laplace equation for the velocity potential, and two boundary conditions: a nonlinear one (kinetic) on the free surface, and zero vertical velocity component at the rigid bottom \cite{3,2016XXX2}. Under the assumption of very small amplitude waves (or steepness) the problem can be considered as a weakly nonlinear one, and the standard way of modeling is to derive the NLS equation by expanding the surface elevation and the velocity potential in power series and using the multiple scale method \cite{2016XXX1,2016XXX2,2018XXX13,3,2018XXX10,dysthe,dysthe2008,bookconf}. 

The procedure is to introduce slow independent
variables (both for time and space) and treat each of them as independent. The extra degrees of freedom arising from such variables allows one to remove the secular terms that may appear in the standard expansion. The multiple scale expansion is usually performed in physical space and a simplification of the procedure is the requirement that the waves are quasi-monochromatic. In the approximation of infinite water depth, for two-dimensional waves the surface elevation, up to third order in nonlinearity, takes the  form 
$$
\eta(x,t)=\biggl( |A(x,t)| -\frac{1}{8} k_{p}^{2} |A(x,t)|^3 \biggr) \cos \theta+\frac{1}{2}k_{p} |A(x,t)|^2 \cos(2\theta)
$$
\begin{equation}\label{eq1}
+\frac{3}{8}k_{p}^{2} |A(x,t)|^3 \cos(3 \theta)+\dots,
\end{equation}
where $A(x,t)$ is a complex wave envelope, $k_p$ is the wave number of the carrier wave, $\eta(x,t)$ is the water elevation, $\theta=(k_p x-\Omega_0 t+\phi)$ is the phase, and $\phi$ a constant phase. In addition we know that $\Omega_0=\omega_p (1+k_{p}^{2} |A(x,t)|^2/2)$ is the nonlinear dispersion relation, with $\omega_p=\sqrt{g k_p}$. The complex envelope obeys the NLS equation
\begin{equation}\label{eq2}
i\biggl( \frac{\partial A}{\partial t} +c_g \frac{\partial A}{\partial x} \biggr) -\frac{\omega_p}{8k_{p}^{2}}\frac{\partial^2 A}{\partial x^2}-\frac{1}{2}\omega_p k_{p}^{2} |A|^2 A=0,
\end{equation}
with $c_g=\partial \omega / \partial k$ being the group velocity. The NLS Eq. (\ref{eq2}) has various types of traveling solutions known as breathers or solitons. 
One analytic solution with major impact in literature is a combine one-parameter $\alpha$ family given by \cite{bookconf}
\begin{equation}\label{eq3in}
A(X,T)=A_0 e^{2 i T}\biggl( 
1+ \frac{2 (1-2 \alpha) \cosh (2 R T)+i R \sinh (2 R T)}{\sqrt{2\alpha} \cos (\Omega X)-\cosh (2 R T)} 
\biggr),
\end{equation}
where the $X,T$ are arbitrary scaled variables by a factor $s$ and the solution $A(x,t)=s A(sX, s^2 T)$, $R=\sqrt{8 \alpha (1-2 \alpha)}$ and $\Omega=2 \sqrt{1-2\alpha}$. When the parameter $\alpha \in (0,0.5)$ Eq. (\ref{eq3in}) describes the space-periodic Akhmediev Breather family (AB), and when $\alpha >0.5$ 
Eq. (\ref{eq3in}) describes the time-periodic Kuznetsov-Ma Soliton (KM) \cite{bookconf,3}. Moreover, in the singular value for parameter $\alpha=0.5$ Eq. (\ref{eq3in}) describes a rational solution known as Peregrine (P) solution \cite{2018XXX20}
\begin{equation}
\label{eq.peregrine}
A(X,T)=A_0 e^{2 i T}\biggl( 
-1+ \frac{4+16 i T}{1+4 X^2+16 T^2} 
\biggr).
\end{equation}
The Peregrine solution in Eq. (\ref{eq.peregrine}) only represents the lowest-order solution of a family of doubly-localized Akhmediev-Peregrine breathers (AP), \cite{2018XXX21,3428,2016XXX4}, also called higher order breathers  \cite{2016XXX7}
\begin{equation}
A_{j}(X,T)=e^{2 i T}\biggl( (-1)^j+\frac{G_j +i H_j}{D_j} \biggr),
\label{doublebreath}
\end{equation}
where the terms $G_j, H_j, D_j$ are polynomials which can be generated by a recursion procedure \cite{3428}.

While in deeper water $3^{rd}$-order nonlinearity causes focusing of long-crested and narrow-banded waves and hence possibility of occurrence of freak waves, in shallower water the nonlinear dynamics are dominated by $2^{nd}$-order nonlinearity. Waves over variable water depth can be modeled for irrotational, inviscid and incompressible flow with a variable coefficient NLS equation. In the approximation of finite depth $(kh)^{-1}=\mathcal{O}(1)$, mild slope $\partial h / \partial x =\mathcal{O}(2)$, and small steepness $\epsilon=\mathcal{O}(3)$ the authors in \cite{5.of.trulsen2012} presented a NLS model with variable coefficients plus a shoaling term.
In this model water surface displacement $\eta$, Eq. \ref{eq1}, and velocity potential $\Phi$ can be written as $3^{rd}$-order perturbation series normalized to $g$ and $\omega_{p}$, respectively
$$
\eta=\epsilon^2 \bar{\eta}+\frac{1}{2}(\epsilon A e^{i \theta}+\epsilon^2 A_{2} e^{2 i \theta}+\dots+\hbox{c.c.}),
$$
\begin{equation}
\label{eq.2012nls.eq5}
\Phi=\epsilon \bar{\phi}+\frac{1}{2}(\epsilon A_{1}^{'} e^{i \theta}+\epsilon^2 A_{2}^{'} e^{2 i \theta}+\dots+\hbox{c.c.}),
\end{equation}
where $\epsilon \theta =\int^{x} k(\xi) d\xi -t$, and c.c. means complex conjugation. The resulting NLS modified (with respect to Eq. \ref{eq2}) equation in terms of the first harmonic amplitude $A$ of the surface displacement is
\begin{equation}
\label{eq.2012nls.eq10}
i\mu \frac{dh}{dx}A+i\biggl( \frac{\partial A}{\partial x}+ \frac{1}{v_g}  \frac{\partial A}{\partial t}\biggr) +\lambda  \frac{\partial^2 A}{\partial t^2}=\nu |A|^2 A,
\end{equation}
where the coefficients $\mu, \lambda, \nu, \bar{\omega}$ depending on $k, h$ and $v_g$ at constant imposed $\omega$  are defined in \cite{5.of.trulsen2012}. In particular, the extra shoaling term  $i \mu h_x$ generalizing the traditional NLS Eq. \ref{eq2} comes from the conservation of wave action flux \cite{with1967}. For the specific bathymetry in our experiments, Figs. \ref{fig1}, \ref{tank}, \ref{fig03}, when the waves travel over the shoal (at $x \sim 11-15$ m from wave-maker) the dispersion coefficient $\lambda(h)$ has only a slow variation of maximum $12\%$ of its value. The nonlinear term coefficient $\nu(h)$ decreases on top of the shoal with $54\%$ of its deep water value, while the shoaling term coefficient $\mu(h)$ has a local increase of $140\%$ on top of the island. The effect of the shoaling term, similar mathematically to the linear dissipative terms occurring in non-homogeneous medium, or to the boundary-layer induced dissipation term in an uniform depth, is a change in wave's amplitude. Actually, it was found \cite{zaxq}, that such damping terms can stabilize the BF instabilities, especially since the nonlinear term contribution decreases in the shoaling regions. This effect is visible in our experiments manifesting as a decrease in the BFI over the shallower region, for any $H_s$ value, Fig. \ref{figBFIvsSpace}.
Analyzing Eq. \ref{eq.2012nls.eq10} with the Djorddjevic\'{c}-Redekopp model \cite{zaxq}, it results that
\begin{equation}
\label{eq.var.bottom.term}
\frac{d h}{d x}\sim -\frac{dk_p}{dx} \sim \frac{d |A|}{dx},
\end{equation}
where these relationships are in effect because the  shoaling term coefficient can be absorbed in the relation $\mu \sim d v_g / dx$. Eq. \ref{eq.var.bottom.term} implies that waves entering in a shallower region experience a decreasing amplitude and wavelength, while waves expanding over deeper regions experience amplitude and wavelength growth. This effect is clearly visible in our results, see for example Figs. \ref{fig03}-\ref{fig02c}, \ref{figxxx}.

In Fig. \ref{figBFIvsSpace} we present the BFI for the three different wave steepness vs. space. Where the water depth is larger (gauges $3\div 8$ and $21\div 28$)  BFI has larger values, and this value increases with the steepness as we can see from the red and blue curves spikes at gauges $7, 22, 27$.  For example, this effect is quite visible over gauges $23\div 28$ where BFI increases monotonically with water depth, and again over gauges $27\div 30$ where BFI decreases monotonically with decreasing of the water depth. Over regions with shallower water depth  ($kh$ is closer to the MI threshold) the BFI decreases no matter of the steepness (see black, red and blue curves over gauges $11\div 20$ in Fig. \ref{figBFIvsSpace}). However, the dynamic response of the waves depends on a combination of water depth (gray curve with circles), bottom slope (green thick curve) and wave steepness (in order of its increasing the upper curves: black, red, blue).  At a sudden drop in the water depth, higher steepness tends to reveal a higher BFI, hence steeper waves are more likely to build RW after shoals and islands (gauge $21$). 

Over regions where water becomes permanently shallower (gauges $30\div 40$) the relaxation distance for decreasing and stabilizing of the BFI, kurtosis and skewness depends on the wave steepness. While at $H_s =3.22$cm the BFI variation is almost monotonically  correlated to the water depth variation, for larger waves with $H_s =5.2 \div 6.2$cm the BFI spikes back to larger values, and is not stabilized for a length of about $8\div 10$ m $\gg \lambda_{p}$ as mentioned in \cite{5.of.trulsen2012}, too.

We also noticed that for small values of the bottom slope in absolute value on the shallow side of the slope, kurtosis and skewness
can stabilize almost at the same location as the change
of depth. Large local values of the absolute value of bottom slope (like fast drops or steep increases of the bottom represented by the spikes of the green curve in Fig. \ref{figBFIvsSpace} over gauges $20-21$ or $31-32$) induce spikes in the BFI and this effect is stronger for larger wave steepness, and less prominent for smoother waves like the case $H_s =3.22$cm. This effect can be correlated  with the  observation of similar spikes in kurtosis and skewness at the same locations, Fig. \ref{figwkusk}, and these observations are in agreement with the experiments in \cite{trulsen2012} and numerical evaluations in \cite{ducro}.  The  increase of skewness and kurtosis over shallowing regions, especially in the transition zone,  was also correlated with deviations of the wave states from the Gaussian distribution and the increase of probability of RW  occurrence. These changes in the statistics parameters of the wave field over transition zone depend on the wave steepness (and consequently on the Ursell number and $H_s$), but not necessary on the length of the transitional zone, as we noticed the occurrence of localized spikes at the beginning of any high bottom slope region which do not necessarily continue along the shallower region.  The results obtained confirm
the conclusion made \cite{25.of.trulsen2012,trulsen2012,truls96}
 in the framework of the nonlinear Schrodinger equation for narrow-banded wind wave field, that kurtosis and the number of
freak waves may significantly differ from the values expected
for a flat bottom of a given depth.

While the wave propagate over the uphill slope, from deeper to shallower water it becomes evidence from Figs. \ref{figwkusk},\ref{figBFIvsSpace} that as long as the shallower side of the slope is sufficiently shallow, and slope length is small enough, we observe local maxima (spikes) of kurtosis, skewness and BFI. These localized maxima are placed at the shallower end of the slopes in agreement with the results from \cite{trulsen2012}. In our experiments the bottom mimics a realistic ocean floor, and the regions with almost constant water depth are not very long, so we do not observe the asymptotic stabilizing of kurtosis and skewness. 


We fit the traveling coherent wave packages obtained din our experiments, see for example the red stripes in Figs. \ref{fig02},\ref{fig02b},\ref{fig02c}, or the wave packages easily visible in Figs. \ref{fig2},\ref{figxxx}, with all the above solutions trying to identify which one describes the best our results.

In Figs. \ref{figbreath} we fit the earliest coherent package formed in small steepness waves with KM solitons. In experiment this group travels as a doublet of stable localized waves, and it is not obvious if this is a bound group of two independent KM solitons, or it is one AB double-breather (higher order Peregrine breather). All theoretical breathers presented Figs. \ref{figbreath} have the same set of parameters, except being translated in space and time accordingly to the gauge position and chosen interval of time. It is very interesting that the match keeps being good enough while the group travels over variable bottom, over a shoal and the deep valley following, and even up the beach when the waves increase in amplitude and become pretty sharp (see the $8^{th}$ frame for example) and ready to break. 

In Figs. \ref{figbreathdouble} we do not show the theory but instead present an overlap of $7$ instants of the same wave group, shifted in time correspondingly. The $8^{th}$ frame shows an obvious match of the same type of behavior for this coherent traveling group, and the likeliness to a breather, possibly a higher-order breather

We also present the match of the stable traveling doublet with two KM solitons bound together, Fig. \ref{fit1} left, as compared to a best fit with a single KM soliton, presented in the right frame. In Figs. \ref{fit2} we fit the experiment with Peregrine breathers (red curves) solitons, and for comparison, the same experimental instants were fitted with KM solitons (blue curves). In Figs. \ref{fit3}, \ref{fit4} we present comparison with double AB breathers, Eq. \ref{doublebreath}. This modeling represents the best match, so we believe that the stable, oscillating and traveling doublets are actually higher order AB submerged in a random wave background. There also a possibility to explain these oscillating and breathing doublets  as Satsuma-Yajima solitons and the supercontinuum  generation effect \cite{3,bookconf}.

Same qualitative results, and the same percentage of matching are obtained for the other two experiments, of higher steepness, but  we do not present them here in detail, in order to keep a reasonable length for the paper. In Figs. \ref{figYYY}, \ref{fig4} we present density plots of the wave heights, in space-time frames, for the steeper waves. These plots show constant group velocity traveling  breathers over the shoal and deep valleys.

In our experiments the mean value of steepness is $0.0765\pm 4\%$, and the theoretical one obtained from the match of experiments with the same KM or peregrine breather results $0.07803$ showing a good match between experiments and theory. The match was made between the analytic form of the KM breather  and the experiment for the gauges $\#4, 10, 12, 18, 19, 20, 23, 25, 28, 30, 34$. Since ocean waves are usually characterized by an average steepness of about $\epsilon \sim 0.1$ corresponding to the peak frequency of the spectrum, both the experimental and theoretical match are plausible. From measuring of the time interval when this structure arrives at various gauges we obtain a group velocity for the breather of $V_g=0.81$ m/s. The theory predicts the occurrence of maximum heights of these breather in the range $A_{max}/A_{0} \sim 3.92$ which is in good agreement with our experimental values of $3.41$.

\section{Conclusions}

In this paper, we present experimental results describing the dynamics of a random background of deep uni-directional, long crested, water waves over a non-uniform bathymetry consisting in  a shoal and several deeper valleys, as well as a final run-up beach. Experiments were performed with waves initially generated with a JONSWAP spectrum, keeping the same carrier (central) frequency, but for three different wave significant heights, involving three different wave steepness. The experimental results confirm the formation of very stable, coherent localized wave packages which travel with almost uniform group velocity across the variable manifolds of the bottom. By using well established statistical tools, and by matching experiments with some of the exact solutions of the NLS equation, we proved that these coherent wave packages coming out of the random background are actually  deep water breathers/solitons solutions (mainly Kuznetsov-Ma, Akhmediev, higher order AB and Peregrine breathers/solitons types), and we put into evidence the formation of rogue waves around those  regions  where the BFI, kurtosis and skewness predict their formation by taking larger values. The evolution and distribution of the statistical parameters, \textit{i.e.}  space and time variation of kurtosis, skewness and BFI, Fourier and moving Fourier spectra, and probability distribution of wave heights, are interpreted in terms of the balance of the terms in a generalized NLS equation for non-uniform bathymetry, having variable coefficients and a shoaling extra term.

\section*{Acknowledgements}
Parts of this work were supported by the Natural Science Foundation
of China under grants NSFC-51639003 and NSFC-51679037. One of
the authors (AL) is grateful to Dalian University of Technology for hospitality during the accomplishment of this research project in 2018-2019. The present work is also supported by National Key Research and Development Program of China (2019 YFC0312400) and  
(2017 YFE0132000). National Natural Science Foundation of China (51975032) and (51939003). The State Key Laboratory of Structural Analysis for Industrial Equipment (S18408).

\end{document}